\documentclass[11pt]{article}
\usepackage{amsfonts}
\usepackage{amssymb}
\usepackage{tikz}
\usepackage{amsmath,amssymb}
\usepackage{bbm}

\def\sfrac#1#2{{\textstyle\frac{#1}{#2}}}
\newcommand{\unit}{\mathbbm{1}}   

\newcommand{\CA}{\mathcal{A}}    

\newcommand{\CF}{\mathcal{F}}

\newcommand{\R}{\mathbb{R}}     
\newcommand{\C}{\mathbb{C}}     
\newcommand{\Hbb}{\mathbb{H}}     
\newcommand{\Nbb}{{\mathbb{N}}}    

\newcommand{\im}{\mathrm{i}} 
 
\newcommand{\dd}{\mathrm{d}}     
\newcommand{\dpar}{\partial}     
\newcommand{\hra}{{\hookrightarrow}}     
\newcommand{\diag}{{\mathrm{diag}}}     

\newcommand{\+}{\dagger}
\newcommand{\wt}{\widetilde}

\newcommand{\Ad}{\mathrm{Ad}} 
\newcommand{\sU}{\mathrm{U}}

\newcommand{\sSO}{\mathrm{SO}}     
\newcommand{\sGL}{\mathrm{GL}}  
 
\newcommand{\ru}{\mathrm{u}}     
 
\newcommand{\rCl}{\mathrm{Cl}}

\newcommand{\al}{{{\alpha}}} 
 \newcommand{\ga}{{{\gamma}}}

\newcommand{\veps}{{\varepsilon}} 
\newcommand{\vph}{{{\varphi}}}

\newcommand{\xh}{\hat{x}} 
\newcommand{\ph}{\hat{p}}

\newcommand{\ab}{{\bar{a}}}
\newcommand{\bb}{{\bar{b}}}

\newcommand{\Av}{{A_{\sf{vac}}}}
\newcommand{\Lv}{{L_{\sf{v}}}}
\newcommand{\qv}{{q_{\sf{v}}}}

\newcommand{\und}{{\quad\mathrm{and}\quad}}

\makeatletter

\@addtoreset{equation}{section}
\makeatother

\begin{document}
\begin{titlepage}
\setcounter{page}{0}
.
\vskip 15mm
\begin{center}
{\LARGE \bf On Dirac equations on phase spaces}\\
\vskip 2cm
{\Large Alexander D. Popov}
\vskip 1cm
{\em Institut f\"{u}r Theoretische Physik,
Leibniz Universit\"{a}t Hannover\\
Appelstra{\ss}e 2, 30167 Hannover, Germany}\\
{Email: alexander.popov@itp.uni-hannover.de}
\vskip 1.1cm
\end{center}
\begin{center}
{\bf Abstract}
\end{center}

\noindent We consider Dirac equations on relativistic phase spaces $T^*{\mathbb R}^{p-1,1}$, where ${\mathbb R}^{p-1,1}$ is Minkowski space with $p=2,4$. We use the geometric quantization approach in which the wave functions are polarized sections of a complex line bundle $L_{\sf{v}}$ over $T^*{\mathbb R}^{p-1,1}$. The covariant derivatives with connection $A_{\sf{vac}}$ in this bundle define canonical commutation relations. Fermions are charged with respect to the field $A_{\sf{vac}}$, so lifting the Dirac equations from space-time ${\mathbb R}^{p-1,1}$ to phase space $T^*{\mathbb R}^{p-1,1}$ results in their solutions being localized in the space ${\mathbb R}^{p-1}$ or in space-time ${\mathbb R}^{p-1,1}$. We describe the explicit form of these solutions.
\end{titlepage}

\newpage
\setcounter{page}{1}

\tableofcontents

\section{Introduction and summary}

\noindent  In this paper we use the differential geometric language introduced into quantum mechanics in the approach of geometric quantization (see e.g. \cite{Sour, Kost, Sni, Wood}). Geometric quantization is a neat mathematical reformulation of the canonical quantization procedure that may improve our understanding of quantum mechanics. In this approach, a complex line bundle $\Lv$ with the structure group U(1)$_{\sf{v}}$ and a connection $\Av$ is introduced on the phase space of the classical system, and the condition of independence of half of the coordinates (the polarization condition) is imposed on sections of the bundle $\Lv$, after which sections can be identified with $\psi$-functions of quantum mechanics. We emphasize that in this paper we are not studying the transition from classical to quantum mechanics, but the geometric aspects of quantum mechanics itself.

To clarify the above, consider the phase space $(T^*{\R}^3, \omega_{\R^6}^{})$ of a non-relativistic particle of mass $m$, where $\omega_{\R^6}^{}=\dd\theta_{\R^6}^{}$ is a symplectic 2-form with potential $\theta_{\R^6}^{}=x^a\dd p_a$ and $x^a, p_a$ are coordinates and momenta, $a=1,2,3$. On a complex line bundle $\Lv$ over $T^*{\R}^3$ we introduce the connection $\Av =-\im\theta_{\R^6}^{}=-\im x^a\dd p_a$ and hence the covariant derivatives\footnote{The dependence of all quantities on Planck's constant $\hbar$  is unimportant for us, so we use the natural units with $\hbar =c=1$.} are
\begin{equation}\label{1.1}
\nabla_a:= \frac{\partial}{\partial x^a}=:\dpar_a  \und
\nabla^{a+3}:=\frac{\partial}{\partial p_a}+\im A^{a+3} =:
\dpar^{a+3}-{\im}\,x^a\ .
\end{equation}
For the curvature of connection $\Av = \im\,A^{a+3}\dd p_a$ we have
\begin{equation}\label{1.2}
F_{\sf{vac}}=\dd \Av = -\im\dd\theta_{\R^6}^{}=-\im\omega_{\R^6}^{}=
\sfrac{\im}2\,F_a^{~b+3}\dd x^a\wedge\dd p_b + \sfrac{\im}2\,F_{~~~~a}^{b+3}\dd p_b\wedge\dd x^a=-\im\,\dd x^a\wedge\dd p_a\ ,
\end{equation}
where components of this background gauge field are 
\begin{equation}\label{1.3}
\im\,F_a^{~b+3}=-\im\,F_{~~~~a}^{b+3}=[\nabla_a, \nabla^{b+3}]=-\im\,\delta_a^b=[\ph_a, \xh^b]
\end{equation}
with operators
\begin{equation}\label{1.4}
\ph_a=-\im\nabla_a=-\im\dpar_a\und \xh^a = \im\nabla^{a+3}=x^a + \im\dpar^{a+3}\ .
\end{equation}
After imposing on sections $\psi$ of the bundle $\Lv\to T^*\R^3$ polarization conditions, 
\begin{equation}\label{1.5}
\dpar^{a+3}\psi = 0\quad\Leftrightarrow\quad\mbox{independence from} \ p_a,
\end{equation}
operators \eqref{1.4} become standard operators of coordinates and momenta. 

The complex line bundle $L_{\C}^+:=\Lv$, introduced in \cite{Sour, Kost, Sni, Wood}, is associated with principal bundle 
$P(T^*\R^3, $ U(1)$_{\sf{v}}$) with structure group U(1)$_{\sf{v}}$. From \eqref{1.1}-\eqref{1.5} we see that the definition of the canonical commutation relations \eqref{1.3} is nothing more than the introduction of a connection $\Av$ and the curvature $F_{\sf{vac}}$ on the bundle $L_{\C}^+$. Both $\Av$ and $F_{\sf{vac}}$ take values in the algebra $\ru(1)_{\sf{v}}=\,$Lie\,U(1)$_{\sf{v}}$.

The structure group U(1)$_{\sf{v}}\cong S^1$ of the bundle $L_{\C}^+$ has a generator $J$ and a charge operator $Q_{\sf{v}}=-\im J$. If we consider $J$ as a derivative with respect to the angle $\theta$ on the circle $S^1$, then sections $\psi$ of the bundle $L_{\C}^+$ will depend on $\theta$ as $e^{\im\theta}$. They are eigenfunctions of the operator $Q_{\sf{v}}=-\im\dpar_\theta ,\  Q_{\sf{v}}\psi=q_{\sf{v}}\psi$, with an eigenvalue $q_{\sf{v}}=1$, which we will call {\it quantum charge}. With the complex line bundle $L_{\C}^+$ one can associate the complex conjugate bundle $L_{\C}^-:=\bar L_{\C}^+$ whose sections depend on $\theta$ as $e^{-\im\theta}$ and have a quantum charge $q_{\sf{v}}=-1$. Sections of bundles $L_{\C}^+$ and $L_{\C}^-$ are associated with particles and antiparticles, respectively.

The field $\Av$ specifies interaction of particles with charge $q_{\sf{v}}$ with vacuum and is associated with the potential energy of particles. Field $\Av$ has nothing to do with electromagnetic interactions and all other fundamental interactions.  For example, the electric charge $q_{\sf{e}}$ is introduced through the defining of complex line bundles $E_\C^+$ for $q_{\sf{e}}>0$ and $E_\C^-$ for $q_{\sf{e}}<0$ over the space $\R^3$ or space-time $\R^{3,1}$ and introducing a connection $A_{\sf{em}}$,
\begin{equation}\label{1.6}
\dpar_t\ \to\ \nabla_t=\dpar_t + \im q_{\sf{e}}A_t\ , \quad \dpar_a\ \to\ \nabla_a=\dpar_a +\im q_{\sf{e}}A_a\ ,
\end{equation}
with components $(A_t, A_a)$ only along the space-time directions. After this, we can introduce four complex line bundles $L_{\C}^+\otimes E_\C^+$, $L_{\C}^+\otimes E_\C^-$, $L_{\C}^-\otimes E_\C^+$, $L_{\C}^-\otimes E_\C^-$ corresponding to four combinations $(q_{\sf{v}}, q_{\sf{e}})=(1, e^+), (1, e^-), (-1, e^+), (-1, e^-)$ of quantum and electric charges. On these bundles, connections with components $(q_{\sf{e}}A_t , \ q_{\sf{e}}A_a , \ 
q_{\sf{v}}A^{a+3})$ will be defined, and the components $(A_t, A_a)$ can be null, but the components  $A^{a+3}$ cannot, they are fixed. 

To clarify the connection between field $\Av$ and the potential energy of interaction of particles with vacuum, we introduce on $T^*\R^3$ the metric
\begin{equation}\label{1.7}
g_{\R^6}=\delta_{ab}\dd x^a\dd x^b + g^{a+3\, b+3}\dd p_a\dd p_b=\delta_{ab}\dd x^a\dd x^b + w^4\delta^{ab}\dd p_a\dd p_b\ ,
\end{equation}
where $w$ is a parameter with the dimension of length (when $\hbar =c=1$), and the covariant Laplacian
\begin{equation}\label{1.8}
\Delta_{\R^6}=\delta^{ab}\dpar_a\dpar_b + g_{a+3\, b+3}\nabla^{a+3}\nabla^{b+3}\ ,
\end{equation}
which on bundles $L_{\C}^\pm$ reduces to the Hamiltonian of quantum harmonic oscillator since
\begin{equation}\label{1.9}
g_{a+3\, b+3}\nabla^{a+3}\nabla^{b+3}=-\frac{1}{w^4}\,\delta_{ab}x^a x^b\ .
\end{equation}
Thus, the first term $\Delta_3=\delta^{ab}\dpar_a\dpar_b$ in \eqref{1.8} specifies the kinetic energy of the particle, and the second term in \eqref{1.8} specifies the potential energy of interaction of a nonrelativistic spin-zero particle with vacuum. Note that in the limit $w^2\to\infty$ the interaction with vacuum is switched off and we get a free particle. 

The inclusion of interaction with $\Av$ leads to the replacement of the Klein-Gordon equation with the Klein-Gordon oscillator equation. For spin $s=1/2$ particles, we will replace the Dirac equation on Minkowski space $\R^{3,1}$, describing free particles, with the Dirac equation on phase space $T^*\R^{3,1}$, describing particles interacting with the vacuum field $\Av$, and describe its solutions. We will show that interaction with the vacuum field $\Av = x^a\dd p_a$ leads to localization of particles in space, and in the case of  $\Av = x^\mu\dd p_\mu$ with $\mu = 0,...,3$, also to localization in time. Discussion of the vacuum background field $\Av$ and the interaction of particles with it is the main topic of this paper.

Finally, we note that interaction with field $\Av$ changes the phase space of particles. Recall that the classical phase space of a free relativistic massive scalar particle is the six-dimensional manifold
\begin{equation}\label{1.10}
X^6 = H^3\times \R^{3,1}/\R^* = (H^3_+\times \R^{3,1}\cup H^3_-\times \R^{3,1})/\R^*\ ,
\end{equation}
where $H^3=H^3_+\cup H^3_-$ is the two-sheet hyperboloid $p^2+m^2=0$ in the momentum space and $\R^*=\sGL(1,\R)$ is a one-parameter Lie group. For a particle interacting with $\Av = x^a\dd p_a$ the above space is changed to the manifold
\begin{equation}\label{1.11}
Y^6 = H^6\times \R/\sGL(1,\R ) = (H^6_+\times \R\cup H^6_-\times \R )/\sGL(1,\R )\ ,
\end{equation}
and for $\Av = x^\mu\dd p_\mu$ we get the simply connected manifold
\begin{equation}\label{1.12}
Z^6 = \Ad S_7/\sU (1)=\sU(3,1)/\sU(3){\times}\sU(1)\ .
\end{equation}
It is clear that interaction with $\Av$ changes the energy-momentum relation, where the contribution of potential energy of interaction with vacuum will appear.

\section{Quantum bundles}

\subsection{Classical particles}

\noindent Let us consider the phase space $T^*\R^3=\R^6$ with coordinates $x^a$ and momenta $p_b$, $a,b = 1,2,3$. The canonical symplectic structure on $T^*\R^3$ is
\begin{equation}\label{2.1}
\omega_{\R^6}=\dd\theta_{\R^6}=\sfrac12\,   \omega_a^{\ \,b+3}\,\dd x^a\wedge\dd p_b+\sfrac12\,\omega^{b+3}_{\quad\ a}\,\dd p_b\wedge\dd x^a=\dd x^a\wedge\dd p_a\ ,
\end{equation}
where 
\begin{equation}\label{2.2}
\theta_{\R^6}=x^a\dd p_a
\end{equation}
is a potential one-form. Classical nonrelativistic particle is a point in $\R^6$ moving along a trajectory $(x^a(t), p_b(t))\in\R^6$ defined by a Hamiltonian vector field
\begin{equation}\label{2.3}
V_H=\omega^a_{\ \,b+3}\,\dpar_a H\dpar^{b+3} + \omega_{b+3}^{\quad\ a}\,\dpar^{b+3}H\dpar_a\quad\mathrm{for}\quad \dpar_a=\frac{\dpar}{\dpar x^a}\und \dpar^{b+3}=\frac{\dpar}{\dpar p_b}\ .
\end{equation}
Here $H=H(x,p)$ is a Hamiltonian function, $(x,p)\in T^*\R^3$. In \eqref{2.3} we used the bivector field
\begin{equation}\label{2.4}
\omega^{-1}_{\R^6}=\omega^a_{\ \,b+3}\,\dpar_a\wedge\dpar^{b+3}= \omega_{b+3}^{\quad\ a}\,\dpar^{b+3}\wedge\dpar_a
\end{equation}
inverse to the two-form $\omega_{\R^6}$ in \eqref{2.1}. 

We consider time-independent Hamiltonians of the form
\begin{equation}\label{2.5}
H(x,p)=T+V=\frac{1}{2m}\,\delta^{ab}p_ap_b + V(x),
\end{equation}
where $T$ is the kinetic energy of a particle, $m$ is its mass and $V(x)$ is its potential energy, $x=(x^a)\in\R^3$. Equations of motion are
\begin{equation}\label{2.6}
\dot{x}^a=\frac{\dd x^a}{\dd t}=V_H x^a \und\dot{p}_a=\frac{\dd p_a}{\dd t}=V_H p_a\ ,
\end{equation}
where $t\in\R$ is an evolution parameter. For Hamiltonian \eqref{2.5}, equations \eqref{2.6} take the form
\begin{equation}\label{2.7}
m\dot{x}^a=\delta^{ab}p_b\und \dot{p}_a=-\dpar_aV=:F_a\ ,
\end{equation}
where $F_a$ is a force acting on the particle. Its action leads to a change in momentum.

On $T^*\R^3$ we will also introduce the metric\footnote{The introduction of a metric on the cotangent bundle $T^*X$ of a manifold $X$ with a metric has been well studied (see e.g. \cite{Yano, Mok}).}
\begin{equation}\label{2.8}
g_{\R^6}=\delta_{ab}\dd x^a\dd x^b + w^4\delta^{ab}\dd p_a\dd p_b
\end{equation}
with the inverse
\begin{equation}\label{2.9}
g_{\R^6}^{-1}=\delta^{ab}\dpar_a\otimes\dpar_b + w^{-4}\delta_{ab}\dpar^{a+3}\otimes\dpar^{b+3}\ ,
\end{equation}
that is we have
\begin{equation}\label{2.10}
g_{ab}=\delta_{ab},\quad g^{{a+3}\ {b+3}}= w^4\delta^{ab}\und g_{{a+3}\ {b+3}}= w^{-4}\delta_{ab}\ .
\end{equation}
Here $w\in \R^+$ is a length parameter (for $\hbar =c=1$) so that $[w^2p_a]=[$length$]=[x^a]$.

\subsection{Quantum particles}

\noindent According to the canonical quantization program \cite{Dirac}, the coordinate functions 
$x^a,\,p_a$ on $\R^6$ are replaced by operators $\xh^a$ and $\ph_a$, acting on a complex 
function $\psi (x,t)$ as
\begin{equation}\label{2.11}
\ph_a\psi (x,t)=-\im\frac{\partial\psi (x,t)}{\partial x^a}\,\und \xh^a\psi (x,t)= x^a\psi (x,t)\ .
\end{equation}
These operators satisfy the canonical commutation relations (CCR):
\begin{equation}\label{2.12}
[\ph_a, \xh^b]=-\im\,\delta_a^b\ .
\end{equation}
The Hamiltonian equations of motion \eqref{2.6} are replaced by the Schr\"odinger equation
\begin{equation}\label{2.13}
\im\,\frac{\partial\psi (x,t)}{\partial t}= H (\xh , \ph ) \psi (x,t)\ .
\end{equation}
Recall that we use the natural units $\hbar =c=1$.

In the geometric quantization approach \cite{Sour}-\cite{Wood} it is shown that $\psi$ in \eqref{2.11} and \eqref{2.13} is a section of the complex line bundle 
\begin{equation}\label{2.14}
\pi\ :\quad L_{\sf v}\ \stackrel{\C}{\longrightarrow}\ T^*\R^3
\end{equation}
over the phase space $T^*\R^3$ with the real polarization condition
\begin{equation}\label{2.15}
\partial^{a+3}\psi = \frac{\partial \psi}{\partial p_a}=0\quad \Rightarrow\quad \psi =\psi (x, t)\ .
\end{equation}
Hence, the operators in \eqref{2.11} are equivalent to the covariant derivatives
\begin{equation}\label{2.16}
\nabla_a:= {\im}\,\ph_a =\dpar_a \und\nabla^{a+3}:=-{\im}\,\xh^a =
\dpar^{a+3}-{\im}\,x^a\ ,
\end{equation}
acting on sections $\psi$ of the bundle \eqref{2.14} subject to condition \eqref{2.15}. 

From \eqref{2.14}-\eqref{2.16} we conclude that in quantum mechanics it was discovered that particles have an additional internal degree of freedom, parametrized by the complex space $\C_u$, attached to each point $u=(x,p)\in\R^6$ of the phase space $\R^6$ of a Newtonian particle. These spaces $\C_u$ are combined into the line bundle \eqref{2.14} which is an extended phase space
\begin{equation}\label{2.17}
L_{\sf v}= \mathop{\bigsqcup}_{u\in\R^6}\C_u = \mathop{\bigcup}_{u\in\R^6}\left\{(u, \psi_u) \mid \psi_u\in\C_u\right\}\ ,
\end{equation}
with the projection \eqref{2.14} onto $T^*\R^3$. A quantum particle is a section $(u, \psi_u)\sim \psi (u,t)$ of the complex line bundle \eqref{2.14} with the polarization condition \eqref{2.15}. Thus, a classical particle is a point $(x(t), p(t))$ in $T^*\R^3$ and a quantum particle is a field $\psi(x,t)$ with values in fibres $\C_{(x,p)}$ of the bundle \eqref{2.14} and these fibres are internal spaces of the particle. The squared modulus of an $L^2$-function $\psi$ is identified with the probability of detecting a particle at point $x\in\R^3$ during the measurement process.

\subsection{Vacuum gauge fields}

In the ``quantum" bundle $\Lv$ (bundle \eqref{2.14} with polarization \eqref{2.15}) a ``quantum" connection $\Av$ (abbreviation ``$\sf v$" and ``$\sf vac$" for vacuum) is given, which specify the parallel transport of polarized sections $\psi$ of $\Lv$, i.e. a way to identify fibres $\C_u$ (internal degrees of freedom) over nearby points $u\in T^*\R^3$. The bundle $L_\C^+:=\Lv$ has the structure group U(1)$_{\sf v}$ and describes particles with quantum charge $q_{\sf v}=1$ which is an eigenvalue of the generator of the group U(1)$_{\sf v}$. In the coordinate representation the connection $\Av$ has the form
\begin{equation}\label{2.18}
\Av = -\im\theta_{\R^6}^{}=-\im\, x^a\dd p_a = \im\, A^{a+3}\dd p_a\ ,
\end{equation}
where $\theta_{\R^6}^{}$ is the potential \eqref{2.2} of the symplectic form \eqref{2.1} on $T^*\R^3$. It is this connection that is specified in covariant derivatives \eqref{2.16}. The curvature of this connection is
\begin{equation}\label{2.19}
F_{\sf vac} = \dd \Av = -\im\,\omega_{\R^6}^{}=-\im\,\dd x^a\wedge\dd p_a\ .
\end{equation}
In components we have
\begin{equation}\label{2.20}
A_a=-\theta_a=0\und A^{a+3}=-\theta^{a+3}=-x^a\ ,
\end{equation}
\begin{equation}\label{2.21}
F_a^{~b+3}=-\omega_a^{~b+3}=-\delta_a^b\und F_{~~~~a}^{b+3}=-\omega^{b+3}_{~~~~a}=\delta_a^b\ .
\end{equation}
Thus, the CCR \eqref{2.12} are equivalent to the equations
\begin{equation}\label{2.22}
\im\, F_a^{~b+3}:=[\nabla_a, \nabla^{b+3}]=-\im\,\delta_a^b
\end{equation}
meaning that the curvature $F_{\sf vac}$ of the connection $\Av$ has constant components \eqref{2.21}.

Note that forms $\theta_{\R^6}^{}$ and $\omega_{\R^6}^{}$ have no sources and define a canonical symplectic structure on the phase space $\R^6$. Therefore, the background gauge field $\Av =-\im \theta_{\R^6}^{}$  and 
$F_{\sf vac} = -\im\,\omega_{\R^6}^{}$ on ${\R^6}^{}$ define the {\it vacuum} of quantum mechanics. It is the constant field $F_{\sf vac}$ that makes the vacuum contribution to the calculation of any quantities. It always arises when using the CCR \eqref{2.22}.

Note also that the partial derivatives $\dpar^{a+3}$ are generators of translations along momenta,
\begin{equation}\label{2.23}
p_a\ \mapsto\ \exp(\veps_{b+3}\dpar^{b+3})\,p_a=p_a+\veps_{a+3}
\end{equation}
and 
\begin{equation}\label{2.24}
\exp(\veps_{a+3}\dpar^{a+3})\,\psi =\psi\quad\mbox{for}\quad \psi\in \Gamma ({\R^6}^{}, L_\C^+)
\end{equation}
due to condition \eqref{2.15}. On the other hand, quantum translations are generated by covariant derivatives\footnote{Note that if we choose the generator $J$ of the group U(1)$_{\sf v}\cong S^1$ as the derivative $\dpar_\theta$ with $\theta\in S^1$, then the covariant derivative $\nabla^{a+3}=\dpar^{a+3}+A^{a+3}\dpar_\theta$ will be a vector field on the total space of the bundle \eqref{2.14}.} $\nabla^{a+3}$, which are the lifting of vector fields $\dpar^{a+3}$  onto the space \eqref{2.14}, so that $\pi_*\nabla^{a+3}=\dpar^{a+3}$. During quantum translations along momenta, the function $\psi$ changes due to U(1)$_{\sf v}$-rotations of the internal space of the particle accompanying the translations \eqref{2.23} and on the infinitesimal level we have
\begin{equation}\label{2.25}
\im\,\delta^{a+3}\psi := \im\,\nabla^{a+3}\psi = \xh^a\psi =x^a\psi\ .
\end{equation}
Thus, the potential energy $V(x)$ and associated force $F_a=-\dpar_a V$ are connected to the vacuum field $\Av$ through quantum translations \eqref{2.25}. Recall that vacuum is defined as a state in which there are no fields of matter (fermions) and no carriers of known fundamental interactions (gauge bosons). Fields $\Av, F_{\sf vac}$ are not included in the above and define a vacuum state in QM. Thus, the field $\Av\in\ru(1)_{\sf v}$ sets the most fundamental force acting on particles due to their interaction with vacuum.

\subsection{Antiparticles and quantum charge}

\noindent {\bf Line bundle $L_\C^-$.} We have introduced a complex line bundle $L_\C^+$ over the phase space $(T^*\R^3, \omega^{}_{\R^6})$ with a connection $\Av$ and curvature $F_{\sf vac}$. In quantum mechanics, polarized sections $\psi (x,t)$ of the bundle $L_\C^+$ are considered to be wave functions describing particles, and it is usually stated that antiparticles appear only in relativistic theory. From a mathematical point of view, this is incorrect. Recall that if $E\to M$ is a complex vector bundle over a manifold $M$ then the complex conjugate bundle $\bar E$ is also defined, obtained by having complex numbers acting through their complex conjugate. In our case, this will be a bundle $L_\C^-:=\bar L_\C^+$ with complex conjugate connection and curvature,
\begin{equation}\label{2.26}
\bar A_{\sf vac} = -\Av = \im\,\theta^{}_{\R^6}\und \bar F_{\sf vac}=-F_{\sf vac}= \im\omega^{}_{\R^6}\ .
\end{equation}
Sections $\phi$ of this bundle describe antiparticles. Comparing \eqref{2.26} with \eqref{2.18}-\eqref{2.22}, we can see that the bundle $L_\C^-\to T^*\R^3$ arises when we quantize the phase space $(T^*\R^3, -\omega^{}_{\R^6})$ with a symplectic structure  that has the opposite sign than in the case of particles. Changing $\omega^{}_{\R^6}\to -\omega^{}_{\R^6}$ is equivalent to changing the sign of time, $t\to -t$, in evolution equations.

\noindent {\bf Complex structure.} To describe the bundles $L_\C^+$ and $L_\C^-$, we need to describe their fibres $\C$ and $\bar\C$. To do this, we consider the vector space $\R^2$ and the matrix
\begin{equation}\label{2.27}
J=(J^A_B)=\begin{pmatrix} 0&-1\\1&0\end{pmatrix} \quad\mathrm{with}\quad J^1_2=-1,\  J^2_1=1
\end{equation}
of complex structure on $\R^2$. Since $J^2=-\unit_2$, eigenvalues of $J$ are $\pm\im$ and it does not have eigenvectors from $\R^2$. However, $J$ can be extended to $\C^2$ and it has eigenvectors from $\C^2$:
\begin{equation}\label{2.28}
v_\pm =\frac{1}{\sqrt{2}}\begin{pmatrix}1\\\mp\im\end{pmatrix}\quad\mathrm{with}\quad
Jv_{\pm}^{}=\pm \im v_\pm\ ,\quad v_\pm^\dagger v_\pm =1\und v_\pm^\dagger v_\mp =0\ .
\end{equation}
Consequently, the space $\C^2$ is decomposed into two orthogonal subspaces,
\begin{equation}\label{2.29}
\C^2=\C\oplus\bar\C , \  \C^2\ni \Psi = \Psi_++\Psi_-=\psi_+ v_+ + \psi_-v_- =\frac{1}{\sqrt 2}\begin{pmatrix}\psi_++\psi_-\\-\im (\psi_+-\psi_-)\end{pmatrix}\ ,
\end{equation}
where  $J\Psi_\pm =\pm\im\Psi_\pm$, $\Psi_+\in\C$ and $\Psi_-\in\bar\C$. The Hermitian metric on this subspaces of $\C^2$ are defined as
\begin{equation}\label{2.30}
\Psi_+^\+\Psi_+ = \psi_+^*\psi_+ =|\psi_+|^2\und
\Psi_-^\+\Psi_- = \psi_-^*\psi_- = |\psi_-|^2    
\end{equation}
and these subspaces are orthogonal since $\Psi_\pm^\+\Psi_\mp =0$.

{\bf Complex bundle $L_{\C^2}$}. Complex line bundles $L_\C^\pm$ are defined as orthogonal subbundles of $\C^2$-vector bundle
\begin{equation}\label{2.31}
L_{\C^2}=L_\C^+\oplus L_\C^-
\end{equation}
with fibres $\C$ and $\bar \C$ as in \eqref{2.29}. Hence, polarized sections of these bundles are vector-functions
\begin{equation}\label{2.32}
\Psi_\pm = \psi_\pm (x,t)v_\pm\in\Gamma (T^*\R^3, L_\C^\pm)
\end{equation}
with the metric \eqref{2.30}. Connection and covariant derivatives on the $\C^2$-bundle \eqref{2.31} have the form
\begin{equation}\label{2.33}
\Av = A^{a+3}J\,\dd p_a\ ,\quad \nabla_a=\dpar_a\und\nabla^{a+3}=\dpar^{a+3}+A^{a+3}J\ ,
\end{equation}
where $J$ from \eqref{2.27} is the generator of the group U(1)$_{\sf v}\cong\sSO(2)_{\sf v}$.
Formulae \eqref{2.33} reduce to \eqref{2.16}, \eqref{2.18}, \eqref{2.22} on $L_\C^+$, and for the bundle $L_\C^-$ one should replace $\im\to -\im$ since $J\Psi_\pm=\pm\im\Psi_\pm$, i.e. $J\to \pm\im$ on $L_\C^\pm$.

{\bf Schr\"odinger equations.} Quantum mechanics considers $\Psi_+\in L_\C^+$ (particles), time evolution of which is described by the Schr\"odinger equation
\begin{equation}\label{2.34}
\im\frac{\dpar\Psi_+(x,t)}{\dpar t} = \hat H_+\Psi_+(x,t)\ \Rightarrow\ \im\frac{\dpar\psi_+(x,t)}{\dpar t} =  H_+\psi_+(x,t)\ .
\end{equation}
Sections $\Psi_-$ of the bundle $L_\C^-$ satisfy the conjugated equation
\begin{equation}\label{2.35}
-\im\frac{\dpar\Psi_-(x,t)}{\dpar t} = \hat H_-\Psi_-(x,t)\ \Rightarrow\ -\im\frac{\dpar\psi_-(x,t)}{\dpar t} = H_-\psi_-(x,t)
\end{equation}
where
\begin{equation}\label{2.36}
\hat H_\pm= \frac{\hat p^2}{2m_\pm} + \hat V(\hat x)\ ,\quad H_\pm= \frac{\hat p^2}{2m_\pm} + V_\pm( x)\und
\hat V(\hat x)v_\pm = V_\pm(x)v_\pm\ ,
\end{equation}
\begin{equation}\label{2.37}
\ph_a = -\im\,\nabla_a=-\im\,\dpar_a\und \xh^a =J\,\nabla^{a+3}=x^a + J\dpar^{a+3}\ ,
\end{equation}
and $Jv_\pm^{}=\pm\im v_\pm^{}$.

Note that equations \eqref{2.34} and \eqref{2.35} are invariant under global transformations of group U(1)$_{\sf v}$ of the form
\begin{equation}\label{2.38}
\Psi_\pm\ \to\ e^{\theta J}\Psi_\pm = (\cos\theta + J\,\sin\theta )\,\Psi_\pm =e^{\pm\im\theta}\Psi_\pm\ ,
\end{equation}
and therefore fields $\Psi_\pm$ have a charge $q_{\sf v}=\pm 1$, which we will call {\it quantum charge}. Recall that charges correspond to the time-invariant generators of the symmetry group that commute with Hamiltonian. In the considered case, the anti-Hermitian generator of the group U(1)$_{\sf v}$ is $J$ and the operator of quantum charge is its Hermitian version
\begin{equation}\label{2.39}
Q_{\sf v} =-\im J=v_+v_+^\+ - v_-v_-^\+=-\sigma_2\ :\quad Q_{\sf v}\Psi_\pm =\pm\Psi_\pm\ ,
\end{equation}
so that $q_{\sf v}=+1$ and $q_{\sf v}=-1$ are eigenvalues of $Q_{\sf v}$ on the bundles $L_\C^+$ (particles) and $L_\C^-$ (antiparticles).

\noindent {\bf Remark.} The charge $Q_{\sf v}$  is related to global transformation \eqref{2.38} preserving the connection \eqref{2.33} on $L_{\C^2}$. Local transformations \eqref{2.38} with $\theta =\theta (x,p)$ define automorphisms of the bundle $L_{\C^2}$. Automorphisms that preserve polarization \eqref{2.15} are gauge transformations, and those that do not preserve polarization are dynamical symmetries. For example, the transformation $\psi\mapsto\exp (\im\,p_ax^a)\psi$ changes the coordinate to momentum representation. From this point of view, $L_\C^\pm$ and $L_{\C^2}$ are examples of framed bundles discussed e.g. in \cite{PopovRMP}.

If we make the background field $\Av$ dynamical by allowing it to be function of $x^a$, then the CCR \eqref{2.22} will be dynamical and this will correspond to back influence of matter on the vacuum via (as yet unknown) equations for the field $F_{\sf vac}$.

\subsection{Quantum charge density}

\noindent The continuity equations for \eqref{2.34} and \eqref{2.35} have the form
\begin{equation}\label{2.40}
\frac{\dpar\rho_\pm}{\dpar t} + \dpar_aj_\pm^a =0\ ,
\end{equation}
where the quantum charge current components,
\begin{equation}\label{2.41}
j_\pm^a =\frac{1}{2m}\,\delta^{ab}(\psi_\pm^*\ph_b\psi_\pm - \psi_\pm\ph_b\psi_\pm^*)\ ,
\end{equation}
have the same form for $\Psi_\pm\in L^\pm_\C$ since $\ph_a$ in \eqref{2.37} is the same for both bundles $L^\pm_\C$. The quantum charge density is
\begin{equation}\label{2.42}
\rho_{\pm} = \Psi_\pm^\+ Q_{\sf v}\Psi_\pm = \pm\psi_\pm^*\psi_\pm =\pm |\psi_\pm|^2\ .
\end{equation}
On the other hand, the probability densities are given by formulae \eqref{2.30} and they are positive for both particles $\Psi_+$ and antiparticles $\Psi_-$.

An important fact is that $\Psi_+$ and $\Psi_-$ are sections of different vector bundles $L_\C^+$ and $L_\C^-$. The sum of these two vector-valued functions is a section $\Psi$ of the bundle \eqref{2.31}. For these sections we have 
\begin{equation}\nonumber
\Psi=\Psi_+(x,t){+}\Psi_-(x,t){=}e^{-\im\omega t}\psi_+(x)v_+{+}e^{\im\omega t}\psi_-(x)v_-
{=}\sfrac12 e^{-\im\omega t}(\psi^1{+}\im\psi^2)\begin{pmatrix}1\\-\im\end{pmatrix}{+} 
\sfrac12 e^{\im\omega t}(\psi^1{-}\im\psi^2)\begin{pmatrix}1\\\im\end{pmatrix}
\end{equation}
\begin{equation}\label{2.43}
=\begin{pmatrix}\cos\omega t&\sin\omega t\\-\sin\omega t&\cos\omega t\end{pmatrix}\begin{pmatrix}\psi^1\\\psi^2\end{pmatrix}=
e^{-\omega tJ}\begin{pmatrix}\psi^1\\\psi^2\end{pmatrix}\sim \begin{pmatrix} e^{-\im\omega t}&0\\0& e^{\im\omega t}\end{pmatrix}\begin{pmatrix}\psi_+\\\psi_-\end{pmatrix}\ ,
\end{equation}
where $e^{\mp\im\omega t}\in\sU(1)_{\sf v}$, $e^{-\omega tJ}\in\sSO(2)_{\sf v}$, $\psi_\pm=\frac{1}{\sqrt{2}}(\psi^1\pm\im\psi^2)$ and $\psi^1$, $\psi^2$ are complex-valued functions of $x\in\R^3$. Formulae \eqref{2.43} of nonrelativistic quantum mechanics show quite clearly why particles and antiparticles in relativistic theories are associated with positive and negative frequencies. The energies of particles and antiparticles in \eqref{2.34} and \eqref{2.35} are non-negative, $E_\pm^{}=\hbar\omega$, and the correspondence principle requires that they be non-negative in relativistic theories as well.

\section{Quantum charge vs electric charge}

\subsection{Quantum charge of  Klein-Gordon field}
\noindent {\bf Nonrelativistic limit of KG.} Note that for the Hamiltonian of a free nonrelativistic particle or antiparticle,
\begin{equation}\label{3.1}
\hat H_0=-\frac{\hbar^2}{2m}\, \delta^{ab}\,\dpar_a\dpar_b\ ,
\end{equation}
both of the equations \eqref{2.34} and \eqref{2.35} can be obtained from the Klein-Gordon (KG) equation\footnote{For greater clarity, when discussing the nonrelativistic limit of the KG equation in this subsection, we will restore the constant $\hbar$ and $c$.}
\begin{equation}\label{3.2}
\left (\frac{1}{c^2}\frac{\dpar^2}{\dpar t^2} - \delta^{ab}\dpar_a\dpar_b + \frac{m^2c^2}{\hbar^2}\right )\phi =0
\end{equation}
when considering the nonrelativistic limit $c\to\infty$ (see e.g. \cite{Greiner1}). Namely, by decomposing the complex scalar field $\phi =\phi_+ +\phi_-$ into positive and negative frequency parts $\phi_+$, $\phi_-$ and redefining them as
\begin{equation}\label{3.3}
\phi_\pm (x, t) = \psi_\pm(x,t)\exp \bigl(\mp\frac{\im mc^2}{\hbar}t\bigr)\ ,
\end{equation}
in the nonrelativistic limit we obtain equations \eqref{2.34} for $\psi_+(x,t)$ and \eqref{2.35} for $\psi_-(x,t)$ with $\hat H_0$ in the right side. This confirms our assertion that \eqref{2.34} and \eqref{2.35} describe particles $\psi_+$ and antiparticles $\psi_-$. 

Returning to the columns $\Psi_+, \Psi_-$ in \eqref{2.34}, \eqref{2.35}, we note that
\begin{equation}\label{3.4}
\Psi = \Psi_++\Psi_-=\frac{1}{\sqrt{2}}\begin{pmatrix}\psi_++\psi_-\\-\im(\psi_+-\psi_-)\end{pmatrix}
\end{equation}
and therefore when in relativistic quantum mechanics  they talk about a scalar function $\phi$ in \eqref{3.2}, they mean only the first (upper) component of the vector-function \eqref{3.4}. However, the correspondence principle requires that when formulating the KG equation \eqref{3.2}, we must take into account that the states with particles and antiparticles in the nonrelativistic limit are orthogonal. In other words, one should consider the KG field $\phi$ not as a scalar, but as a $\C^2$-valued section \eqref{3.4} of the quantum bundle \eqref{2.31} and similarly with the Dirac field, since the nonrelativistic limit of the Dirac equation also leads to independent equations \eqref{2.34} and \eqref{2.35} (via the Pauli equations).

\noindent {\bf Quantum charge $q_{\sf v}$}. Recall that in the Klein-Gordon theory the scalar product of two complex functions $\phi$ and $\chi$ is introduced with formula \cite{Greiner2}
\begin{equation}\label{3.5}
(\phi , \chi)=\im\int\dd^3x\,\bigl(\phi^*(x,t)\dot\chi (x,t)-\dot\phi^*(x,t)\chi (x,t)\bigr)\ ,
\end{equation}
where dot means $\dpar/\dpar t$. The KG Lagrangian is invariant under the transformation  $\phi\to e^{\im\theta}\phi$  and this relates \eqref{3.5} to the conserved charge
\begin{equation}\label{3.6}
q_{\sf v}=\int\dd^3x\,j^0(x)   =  \frac{\im\hbar}{2mc^2} \int\dd^3x\,  (\phi^*\dot\phi-\dot\phi^*\phi) = \frac{\hbar}{2mc^2}\,(\phi , \phi)\ ,
\end{equation}
which, in nonrelativistic approximation \eqref{3.3}, reduce for $c\to\infty$ to
\begin{equation}\label{3.7}
q_{\sf v}=\frac{\hbar}{2mc^2}\,(\phi , \phi)\quad\to\quad q_{\sf v}= \int\dd^3x\,  (\psi^*_+\psi_+ -\psi^*_-\psi_-)\ .
\end{equation}
Thus, we conclude that for normalized functions $\psi_\pm$ we get the quantum charges
\begin{equation}\label{3.8} 
q_{\sf v}=\int\dd^3x\,   \rho_\pm =\pm \int\dd^3x\, \psi_\pm^*\psi_\pm\ ,
\end{equation}
 i.e. $q_{\sf v}=+1$ for particles $\psi_+$ and $q_{\sf v}=-1$ for antiparticles $\psi_-$.

\noindent{\bf Quantum charge density.} In terms of sections \eqref{3.4} of $\C^2$-bundle \eqref{2.31} the scalar product \eqref{3.5} in the nonrelativistic limit is reduced to the integral \eqref{3.7} with the charge density
\begin{equation}\label{3.9}
\rho =\Psi^\+ Q_{\sf v}\Psi = \psi_+^*\psi_+ - \psi_-^*\psi_- =\rho_+ + \rho_-\ ,
\end{equation}
where $Q_{\sf v}$ is the quantum charge operator, $Q_{\sf v}=-\im J$. Note that when lifting to the KG equation for $\Psi\in L_{\C^2} = L_\C^+\oplus L_\C^-$, the matrix $Q_{\sf v}$ goes over to $\im\dpar_t$, and in the case of the Dirac field $\Psi$ with values in $L_{\C^2}$, the quantum charge density $\bar\Psi\gamma^0\Psi =\Psi^\+\Psi$ cannot be reduced at $c\to\infty$ to \eqref{3.9} because it does not contain a time derivative. The correspondence principle requires changing the definition of Dirac conjugation as $\bar\Psi^{q_{\sf v}}=\Psi^\+\gamma^0\otimes
Q_{\sf v}$, which gives the correct form \eqref{3.9} of the quantum charge density for the limit of $\rho =\bar\Psi^{q_{\sf v}}\gamma^0\Psi$.

\subsection{Vacuum force: summary}

From a geometric point of view, canonical quantization of Newtonian particle is equivalent to replacing a point $(x(t), p(t))$ in the phase space $(T^*\R^3, \omega^{}_{\R^6})$ of particle with a function $\psi_+(x,t)\in \Gamma (T^*\R^3, L_\C^+)$ that specifies the quantum charge distribution density $\rho_+=\psi^*_+\psi_+$ as well as specifying a new type of Abelian gauge field $\Av\in\ru(1)_{\sf v}$ defined on U(1)$_{\sf v}$-bundle $L_\C^+$ over $T^*\R^3$. The probabilistic interpretation of quantum charge density modulus is secondary; the probability of finding a quantum particle is greater where the quantum charge density is higher, and the limit $\psi (x)\to\delta (x)$ is the exact localization of the charge at $x\in\R^3$. A quantum particle is a point of phase space with an internal space $\C$ (fibres of the bundle $L_\C^+$), which is acted upon by a field $\Av$. Thus, a quantum particle is a field 
$\psi_+(x,t)\in L_\C^+$ and it is this field that is primary, a classical particle is an approximation of a quantum one, when the internal space of the particle is not taken into account (shrinks to a point, 
U(1)$_{\sf v}\cong S^1\to 1$, when $\hbar\to 0$).

The fields $\Av$ and $F_{\sf vac}=\dd\Av$, connection and curvature on the quantum bundle $L_\C^+\to\R^6$, are not dynamical, they are given by formulae \eqref{2.18}-\eqref{2.21} encoding the symplectic geometry of the phase space $(\R^6, \omega_{\R^6}^{})$. The components of the curvature $F_{\sf vac}$ are proportional to the components of the ``flat" symplectic 2-form  $\omega_{\R^6}^{}$. Hence, the fields $(\Av, F_{\sf vac})$ should be considered as vacuum fields since the 2-form $\omega_{\R^6}^{}$ has no sources, it is a flat geometric structure on 
$\R^6$, similar to how Minkowski metric defines a flat space $\R^{3,1}$ which is a canonical vacuum in the theory of gravity. Thereby, $(\Av, F_{\sf vac})$ are background fields, they have no sources and define vacuum state of quantum mechanics (QM). Thus, nonrelativistic QM introduces the internal space $(\R^2, J)\cong\C$ of particles, fibres of the bundle $L_\C^+$, and gauge field $\Av$ acting on it, defining interaction of quantum particles with vacuum.

Antiparticles have the phase space $(\R^6, -\omega_{\R^6}^{})$ with the sign of the symplectic structure reversed, which is equivalent to replacing $t\to -t$ in the Hamiltonian equations of motion. Their internal space is $(\R^2, -J)\cong\bar\C$ which leads to the replacement of the bundle $L_\C^+$ by the complex conjugate bundle $L_\C^-$ with the opposite quantum charge, $q_{\sf v}=1\ \to\ q_{\sf v}=-1$.

\subsection{Electric charge}
\noindent Bundles $L_\C^\pm$ with structure group U(1)$_{\sf v}$ are associated with quantum charges $q_{\sf v}=\pm 1$. The connection and curvature on these bundles have the form
\begin{equation}\label{3.10}
(q_{\sf v}\Av , q_{\sf v}F_{\sf vac})=-\im (q_{\sf v}\theta_{\R^6}^{}, q_{\sf v}\omega_{\R^6}^{})\ ,
\end{equation}
where $\theta_{\R^6}^{}$ and $\omega_{\R^6}^{}$ are given in \eqref{2.1} and \eqref{2.2}. It is well known that electric charge $q_e=\pm e$ of particles and antiparticles can be introduced by replacing
\begin{equation}\label{3.11}
p_a\ \to\ p_a +q_e A_a(x)\ ,
\end{equation}
which leads to {\it deformation} of $\theta_{\R^6}^{}$ and $\omega_{\R^6}^{}$ of the form
\begin{equation}\label{3.12}
\theta_{\R^6}^{}\ \to\ \theta_{\R^6}^{e}=\theta_{\R^6}^{}-q_eA
\quad\mbox{for}\quad A=A_a\dd x^a\ ,
\end{equation}
\begin{equation}\label{3.13}
\omega_{\R^6}^{}\ \to\ \omega_{\R^6}^e = \omega_{\R^6}-q_eF = \omega_{\R^6}-q_e\dd A   \ ,
\end{equation}
where in \eqref{3.12} we discarded the term $\dd (q_ex^aA_a)$ since $\dd^2=0$. Accordingly, after quantization we get the fields
\begin{equation}\label{3.14}
\CA_{q_{\sf v}q_e}^{}=q_e A_{\sf em} + q_{\sf v}\Av = \im q_e A - \im q_{\sf v}\theta_{\R^6}^{}\ ,
\end{equation}
\begin{equation}\label{3.15}
\CF_{q_{\sf v}q_e}^{}=q_e F_{\sf em} +q_{\sf v}F_{\sf vac}=\im q_e \dd A - \im q_{\sf v}\omega_{\R^6}^{}\ .
\end{equation}
Note that in 
\begin{equation}\label{3.16}
\CA_{q_{\sf v}q_e}^{}=\im q_e A_a\dd x^a + \im q_{\sf v} A^{a+3}\dd p_a
\end{equation}
the components $A^{a+3}=-x^a$ are fixed by CCR, and the magnetic potential $A_a\dd x^a$ is dynamical and can be equal to zero. By this we want to emphasize that $\Av$ is a vacuum, and $A_{\sf em}$ is an excitation above the vacuum.

Deformation of the symplectic structure \eqref{3.11}-\eqref{3.13} on $T^*\R^3$ leads to a change in the covariant derivatives \eqref{2.16}, 
\begin{equation}\label{3.17}
\partial_a\to\nabla_a=\dpar_a+\im q_eA_a\ ,\quad \nabla^{a+3}=\dpar^{a+3} + \im q_{\sf v} A^{a+3}\ .
\end{equation}
Formulae \eqref{3.17} are often supplemented with the replacement
\begin{equation}\label{3.18}
\dpar_t\ \to\ \nabla_t = \dpar_t + \im q_e A_t\ ,
\end{equation}
with the electric potential $A_t$, as for example in the nonrelativistic Pauli equation.

\subsection{Quantum and electric charge combinations}

\noindent {\bf Tensor products of bundles.} From \eqref{3.14}-\eqref{3.16} it follows that connections $\CA_{q_{\sf v}q_e}^{}$ are defined on four different bundles corresponding to four possible combinations of quantum charge $q_{\sf v}$ and electric charge $q_e$:
\begin{equation}\label{3.19}
(q_{\sf v}, q_e) = (1, e), (1, -e), (-1, e)\und (-1, -e)\ .
\end{equation}
They are described as follows. Let us introduce bundles $E_\C^\pm\to\R^3$ over the coordinate space $\R^3$ with connection $q_e A_{\sf em}$ and gauge group U(1)$_{\sf em}$. Then we pullback bundles $E_\C^\pm$ to  $T^*\R^3$ using the projection $T^*\R^3\to\R^3$ and denote them with the same letters $E_\C^\pm$. After that we introduce the tensor product of bundles, 
\begin{equation}\label{3.20}
(L^+_\C\oplus L^-_\C)\otimes (E^+_\C\oplus E^-_\C) = L^+_\C\otimes E^+_\C \oplus L^+_\C\otimes E^-_\C\oplus L^-_\C\otimes E^+_\C\oplus L^-_\C\otimes E^-_\C\ ,
\end{equation}
whose sections are the vectors
\begin{equation}\label{3.21}
\Psi = \psi_{++}\,v_+\otimes v_+ + \psi_{+-}\,v_+\otimes v_- +\psi_{-+}\,v_-\otimes v_+ +\psi_{--}\,v_-\otimes v_-
\end{equation}
taking values in the space $\C^4$ of fibres of the vector bundle \eqref{3.20}. Note that the tensor products of basis vectors can be constructed as follows:
\begin{equation}\label{3.22}
v_\pm{=}\frac{1}{\sqrt 2}\begin{pmatrix}1\\\mp\im\end{pmatrix}\ \Rightarrow\ v_+{\otimes}v_+{=}\frac{1}{\sqrt 2}
\begin{pmatrix}v_+\\-\im v_+\end{pmatrix}{=}\frac{1}{ 2}\begin{pmatrix}1\\-\im\\-\im\\-1\end{pmatrix},
\ v_+{\otimes}v_- {=}\frac{1}{\sqrt 2}\begin{pmatrix}v_+\\\im v_+\end{pmatrix}{=}\frac{1}{2}\begin{pmatrix}1\\-\im\\\im\\1\end{pmatrix} \mathrm{etc.}
\end{equation}
Thus, we have four line bundles in \eqref{3.20} with four connections \eqref{3.16} on them and four curvature tensors 
\eqref{3.15} for $\CA_{q_{\sf v}, q_e}$. 

\noindent{\bf Generators of U(1)$_{\sf v}\times$U(1)$_{\sf em}$.} It may be more convenient to use $\C^4$-bundle
\eqref{3.20} instead of its $\C$-subbundles. On vectors \eqref{3.21}, the action of the group  U(1)$_{\sf v}$ (first index $\pm$) and the group U(1)$_{\sf em}$ (second index $\pm$) is given, and the generators of the group U(1)$_{\sf v}\times$U(1)$_{\sf em}$ are $4\times 4$ matrices
\begin{equation}\label{3.23}
I_{\sf v} = J\otimes\unit_2\und I_e = \unit_2\otimes J\ ,
\end{equation}
where $J=-\im\sigma_2$. Note that we have two pairs of complex conjugate bundles,
\begin{equation}\label{3.24}
L^+_\C\otimes E^-_\C \und L^-_\C\otimes E^+_\C\ ,\quad
L^+_\C\otimes E^+_\C \und L^-_\C\otimes E^-_\C\ ,
\end{equation}
with sections $\psi_{+-}$ and $\psi_{-+}$, $\psi_{++}$ and $\psi_{--}$. Indices $\pm$ of these vectors indicate the combinations of charges $q_{\sf v}=\pm 1$ and $q_e=\pm e$, where $e>0$ is the modulus of the electric charge of electron. Therefore, particles and antiparticles always belong to either the first or the second pair of bundles \eqref{3.24}. From \eqref{3.23} we see that operators of quantum and electric charges are
\begin{equation}\label{3.25}
Q_{\sf v} = -\im J\otimes\unit_2\und Q_e=-\im e \unit_2\otimes J\ .
\end{equation}
These charges corresponds to groups U(1)$_{\sf v}$ and U(1)$_{\sf em}$. 

\noindent{\bf Charge densities.} Note that if we redesignate the components of the vector \eqref{3.21} as $\psi_{q_{\sf v}q_e}^{}$ with $(q_{\sf v}, q_e)$-data from \eqref{3.19}, then the $L^2$-functions
\begin{equation}\label{3.26}
\rho_{q_{\sf v}q_e}^{}=\psi_{q_{\sf v}q_e}^{*}\psi_{q_{\sf v}q_e}^{}
\end{equation}
will specify the probability density of finding a particle with charges $(q_{\sf v}, q_e)$ at point $x\in\R^3$, and functions
\begin{equation}\label{3.27}
\rho_{q_{\sf v}}^{}=q_{\sf v}\rho_{q_{\sf v}q_e}^{}\und \rho_{q_e}^{}=q_e\rho_{q_{\sf v}q_e}^{}
\end{equation}
will specify the quantum and electric charge densities.

\noindent{\bf Zero charge condition.} It is generally accepted that quarks and antiquarks have fractional electric charge $q_e$ equal $\pm\sfrac13\,e$ or $\pm\sfrac23\,e$. For generator $J$ of the group U(1)$_{\sf em}=S^1$ implemented as a derivative $J=\dpar_\vph$ with respect to the angle $\vph$ on $S^1$, this means that quarks depend on $\vph$ as $\exp (\im q_e\vph )$ with $q_e=\pm\sfrac13\,e,\ \pm\sfrac23\,e$. By introducing basis vectors $v_\pm^{\pm 1/3}$ we can introduce bundles $E^{\pm 1/3}_\C$ with $q_e=\pm\sfrac13\,e$ and obtain complex line bundles $E^{q_e}_\C$ for any charge $q_e$ by taking tensor products of the bundles $E^{\pm 1/3}_\C$. Similarly, we introduce quantum bundles $L_\C^{q_{\sf v}}$ with quantum charges $q_{\sf v}=\pm\sfrac13, \pm\sfrac23$ and $\pm 1$.

It is believed that the total electric charge of the universe is zero. If we assume that the electric charge is related to the quantum charge by the formula
\begin{equation}\label{3.28}
q_e = e^-\qv = -e\qv\ ,
\end{equation}
then the total quantum charge of the universe will also be equal to zero. In this case, only the first two complex conjugate bundles in \eqref{3.24} should be considered.

\section{Fermions}

\subsection{Covariant Laplacian}

In the previous sections we introduced and discussed quantum bundles $L_\C^{\qv}$ and connections on them. Everything related to electromagnetic fields in quantum mechanics is well studied, so we will focus on bundles $L_\C^\pm$ with $\qv =\pm 1$ and connection $\Av$ on them.

Note that in the relativistic case, the matter fields take values in the bundle $L_{\C^2}^{}=L_\C^+\oplus L_\C^-$. Covariant derivatives in this bundle have the form
\begin{equation}\label{4.1}
\nabla_a =\dpar_a\und \nabla^{a+3}=\dpar^{a+3}+ A^{a+3}J =\dpar^{a+3} - x^aJ\ ,
\end{equation}
where $J$ is the generator  of the group U(1)$_{\sf v}$, $Jv_\pm = \pm\im v_\pm$. Recall that the explicit form of the vacuum field $\Av$ is fixed by the CCR.

We introduce covariant Laplacian on the space $T^*\R^3$ in the standard way as the operator\begin{equation}\label{4.2}
\Delta^{}_{6}:=\delta^{ab}\nabla_a\nabla_b + g_{a+3\, b+3}\nabla^{a+3}\nabla^{b+3}\ .
\end{equation}
It is easy to see that on polarized sections $\Psi =\Psi_++\Psi_-$ of the bundle $L_{\C^2}$ operator \eqref{4.2} takes the form
\begin{equation}\label{4.3}
\Delta^{}_{6}\Psi=\bigl (\delta^{ab}\dpar_a\dpar_b - \frac{1}{w^4}\delta_{ab}x^ax^b\bigr)\Psi\ .
\end{equation}
From \eqref{4.3} it follows that $\Delta_6$ is a scalar operator on sections $\Psi_\pm\in L_\C^\pm$ coinciding with the Hamiltonian of the quantum isotropic harmonic oscillator,
\begin{equation}\label{4.4}
H_{\sf osc}=-\frac{1}{2m}\,\Delta_6=-\frac{1}{2m}\,\delta^{ab}\dpar_a\dpar_b + \frac{m\omega^2}{2}\delta_{ab}x^ax^b
\end{equation}
if we put
\begin{equation}\label{4.5}
\frac{1}{w^4}=m^2\omega^2\ .
\end{equation}
Thus, the field $\Av$ included in the covariant Laplacian \eqref{4.2} is associated with the energy
\begin{equation}\label{4.6}
V(x)=-\frac{1}{2m}\,g_{a+3\, b+3}\nabla^{a+3}\nabla^{b+3}=\frac{m\omega^2}{2}\delta_{ab}x^ax^b\ ,
\end{equation}
which we identify with the potential energy of a particle interacting with vacuum. Note that this energy tends to zero as $\omega\to 0$ ($w^2\to\infty$) and in this limit Hamiltonian \eqref{4.4}  tends to the Hamiltonian
\begin{equation}\label{4.7}
H_0=T=-\frac{1}{2m}\,\Delta_3=-\frac{1}{2m}\,\delta^{ab}\dpar_a\dpar_b
\end{equation}
of a free particle. From \eqref{4.2}-\eqref{4.6} it follows that vacuum behaves like an elastic medium.

\subsection{Spinors}

{\bf Metric on $T^*\R^3$}. Let us now discuss the introduction of spin degrees of freedom of particles. A nonrelativistic particle has phase space $\R^6$ with symplectic structure \eqref{2.1} and metric \eqref{2.8}. Note that the components $g_{a+3\, b+3}$ of the metric in \eqref{2.8}-\eqref{2.10} have dimensions $L^4$ since $[p_a]=L^{-1}$ and $[x^a]=L$. These components can be made dimensionless by introducing a fixed length $w_0$ in \eqref{2.8},
\begin{equation}\label{4.8}
g_{\R^6}=\delta_{ab}\dd x^a\dd x^b + \biggl(\frac{w}{w_0}\biggr)^4\delta^{ab}\dd (w_0^2p_a)\dd (w_0^2p_b)\ ,
\end{equation}
and use $\tilde p_a=w_0^2p_a$ with the dimension of length and dimensionless parameter $w/w_0$ in all formulae. We will not move from formula \eqref{2.8} to \eqref{4.8} to make it easier to keep track of the dimensions of different quantities.

\noindent {\bf Symplectic structure on superspace.} Tangent space to $\R^6$ has the vector fields $\dpar_a$, $\dpar^{a+3}$ as its basis. Any vector $\xi$ of the tangent space has the form
\begin{equation}\label{4.9}
\xi =\xi^a\dpar_a + \xi_{a+3}\dpar^{a+3}\ ,
\end{equation}
where for the dimensions of the vector components we have $[\xi^a]=L^0$ and $[\xi_{a+3}]=L^{-2}$. If we take as 
$\xi^a$ and $\xi_{a+3}$ the generators of the Grassmann algebra $\Lambda (\R^6)$, then the Grassmann-valued vectors \eqref{4.9} will be elements of the space $\Pi\R^6$, where the operator $\Pi$ inverts the Grassmann parity of components \cite{Kost}. The Euclidean metric \eqref{2.8} is a symplectic structure on $\Pi\R^6$, so the symplectic structure on $\R^6\times\Pi\R^6$ is \cite{Kost}
\begin{equation}\label{4.10}
\Omega =\omega^{}_{\R^6} + \omega^{}_{\Pi\R^6}=\dd x^a\wedge\dd p_a + \delta_{ab}\,\dd\xi^a\dd\xi^b + g^{a+3\, b+3}\dd\xi_{a+3}\dd\xi_{b+3}\ .
\end{equation}
When quantizing the space $\R^6\times\Pi\R^6$, we obtain differential operators \eqref{2.16} for the bosonic part and first order operators on $\Pi\R^6$,
\begin{equation}\label{4.11}
\xi^a\to\Gamma^a\ ,\quad \xi_{a+3}\to \Gamma_{a+3}\quad\mbox{with}\quad\bigl\{\Gamma^a,\Gamma^b\bigr\}=2\delta^{ab}\und\bigl\{\Gamma_{a+3},\Gamma_{b+3}\bigr\}=2g_{a+3\,b+3}\ ,
\end{equation}
for the fermionic part.

\noindent {\bf Clifford algebra and spinors.} Operators \eqref{4.11} are generators of the Clifford algebra Cl(6,0) and its complexification Cl$^\C$(6)=Cl(6,0)$\otimes\C$. Let us also introduce operators
\begin{equation}\label{4.12}
\Upsilon^a=\sfrac12\,\bigl(\Gamma^a - \im w^2\delta^{ac}\Gamma_{c+3}\bigr)\und 
\Upsilon^{\bar b}=\sfrac12\,\bigl(\Gamma^b+ \im w^2\delta^{bd}\Gamma_{d+3}\bigr)
\end{equation}
satisfying the equations
\begin{equation}\label{4.13}
\{\Upsilon^a, \Upsilon^b\}=0\ ,\quad \{\Upsilon^\ab, \Upsilon^\bb\}=0\und \{\Upsilon^a, \Upsilon^\bb\}=\delta^{a\bb}\ .
\end{equation}
From \eqref{4.13}  we see that the operators $\Upsilon^a$ and $\Upsilon^\ab$ are generators of two Grassmann subalgebras $\Lambda (\C^3)$ and $\Lambda (\bar\C^3)$ in the Clifford algebra Cl$^\C$(6). 

When quantizing the space $\R^6\times\Pi\R^6$, it is necessary to impose on functions on this graded space the condition of dependence on only half of the variables (polarization) \cite{Kost}, condition \eqref{2.15} for the bosonic part and condition $\Upsilon^a\psi =0$ for the fermionic part. After this, functions on $\R^3$ with values in the spinor space are introduced as the minimal left ideal
\begin{equation}\label{4.14}
\mbox{Cl}^\C(6)|0\rangle =\Lambda (\bar\C^3)|0\rangle\cong\C^8\ ,
\end{equation}
where the fermionic vacuum $|0\rangle=\Upsilon^1\Upsilon^2\Upsilon^3$ is the primitive nilpotent element in the algebra Cl$^\C(6)$. Thus, the vector space of spinors \eqref{4.14} on $\R^6$ is isomorphic to the Grassmann algebra 
$\Lambda (\bar\C^3)\subset\mbox{Cl}^\C(6)$ of the subspace $\bar\C^3$ in the space $\R^6\otimes\C=\C^3\oplus\bar\C^3$ and spinors can be considered as functions on the space $\Pi\bar\C^3$ with differential operators \eqref{4.12} acting on them.\footnote{The explicit form of these operators is not important here.} In a similar way, spinors can be introduced on spaces of any dimension and signature. 

\noindent {\bf Matrix representations.} Note that in the finite-dimensional case (unlike QFT) it is more convenient to use matrix representations of generators \eqref{4.11} of the Clifford algebra\footnote{A brief overview of matrix representations of Clifford algebras Cl($p,q$) of spaces $\R^{p,q}$ with metric of signature ($p,q$) can be found in \cite{Coq}.} using the isomorphism
\begin{equation}\label{4.15}
\mbox{Cl}^\C(6)\cong \mbox{Mat}(8, \C)\ .
\end{equation}
Then the spinors are given by columns $V=\C^8$ with left action of the matrix algebra  \eqref{4.15}. Note that formula  \eqref{4.14} defines this column as a left ideal in Cl$^\C(6)$  and a nilpotent element $|0\rangle $ (vacuum) can always be chosen so that this ideal coincides with the last column in Mat(8, $\C$). These are standard constructions. 

\subsection{Pauli and Dirac equations}

\noindent {\bf Two Laplacians.} Returning to the consideration of bundles $L_\C^\pm$, we note that we have two canonical differential operators on $T^*\R^3$:
\begin{equation}\label{4.16}
\mbox{Laplace\ operator}\quad\Delta_3=\delta^{ab}\dpar_a\dpar_b
\end{equation}
and
\begin{equation}\label{4.17}
\mbox{Laplace\ operator}\quad\Delta_6=\Delta_3+g_{a+3\,b+3}\nabla^{a+3}\nabla^{b+3}\ .
\end{equation}
Recall that \eqref{4.16} defines the kinetic energy $T$ of particle, $\Delta_3=-2mT$, and $\Delta_6-\Delta_3=-2mV(x)$ defines the potential energy of particle in QM-vacuum, so that $\Delta_6$ is proportional to the full energy $H_{\sf osc} =T+V$.

\noindent {\bf Pauli equation.} Pauli used the operator $\Delta_3$ of the kinetic energy of a nonrelativistic particle to generalize the description of electron in hydrogen atom. Namely, he introduced a ``square root" of the operator $\Delta_3$,
\begin{equation}\label{4.18}
\Delta_3^{1/2}=\sigma^a\dpar_a\quad\Rightarrow\quad (\sigma^a\dpar_a)^2=\Delta_3\ ,
\end{equation}
thereby introducing spinors $\psi\in\C^2$ for a real Clifford algebra Cl(3,0)$\,\cong\,$Mat(2,$\C$) with Pauli matrices $\sigma^a$ as generators. After that, he introduced the electromagnetic field according to formulae \eqref{3.17}, \eqref{3.18} and wrote out the equation
\begin{equation}\label{4.19}
\im (\dpar_t + \im q_eA_t)\psi = -\frac{1}{2m}\,\bigl(\sigma^a(\dpar_a+\im q_eA_a)\bigr)^2\psi\ ,
\end{equation}
where the potential energy of the electron $\psi$ in the field of proton is introduced through the component $A_t$. Note that Pauli introduced a matrix representation for the real Clifford algebra Cl(3,0), and its complexification is given by the reducible representation
\begin{equation}\label{4.20}
\mbox{Cl}^\C(3)\cong \mbox{Mat}(2, \C)\oplus\mbox{Mat}(2, \C)
\end{equation}
describing bispinors $\Psi\in\C^2\oplus\C^2$.

\noindent {\bf Dirac equation.} Dirac generalized Pauli's logic by taking ``square root" of the d'Alembert operator $\square_4=-\dpar_t^2 +\Delta_3$ associated with the kinetic energy of a relativistic particle,
\begin{equation}\label{4.21}
\square^{1/2}_4=\im\ga^\mu\dpar_\mu\quad\Rightarrow\quad\bigl(\im\ga^\mu\dpar_\mu\bigr)^2=\square_4\ ,
\end{equation}
where $\ga^\mu$ with $\mu =0,...,3$ are $4\times 4$ Dirac matrices which are generators of the Clifford algebra
\begin{equation}\label{4.22}
\mbox{Cl}^\C(4)\cong \mbox{Mat}(4, \C)\ .
\end{equation}
Using formulae \eqref{3.17}, \eqref{3.18} again, we obtain the Dirac equation for an electrically charged particles of spin $s=1/2$,
\begin{equation}\label{4.23}
\im\gamma^\mu (\dpar_\mu + \im q_eA_\mu )\psi - m\psi =0\ ,
\end{equation}
where $\psi\in\C^4$ is a four-component spinor.

\subsection{Extended Klein-Gordon and Dirac equations}

The purpose of this paper is to consider interaction of particles with vacuum described by field $\Av$, which is given by the canonical commutation relations \eqref{2.12} and \eqref{2.22}. To introduce the interaction of a nonrelativistic particle with this field, one should take the ``square root" of the operator $\Delta_6$ and not the operator $\Delta_3$. In other words, one should introduce the Dirac operator
\begin{equation}\label{4.24}
\Gamma_{\R^6}^{}=\Gamma^a\nabla_a + \Gamma_{a+3}\nabla^{a+3}
\end{equation}
on the phase space $T^*\R^3$, where the generators \eqref{4.11} of the Clifford algebra Cl$^\C$(6) are taken as 8$\times$8 matrices, and the covariant derivatives  $\nabla_a$, $\nabla^{a+3}$ are defined in \eqref{4.1}. Operator \eqref{4.24} acts on spinors $\Psi^A\in\C^8$, where $A=1,2$ is the index parametrizing fibres of the bundle $L_{\C^2}$. In other words, $\Psi=(\Psi^A)$ take values in $\C^8\otimes\C^2=\C^{16}$. 

Multiplying the operator  \eqref{4.24} by itself we get
\begin{equation}\label{4.25}
\Gamma_{\R^6}^{2}=\Delta_6 + {\im}\,[\Gamma^a, \Gamma_{a+3}]\ ,
\end{equation}
where the last term in \eqref{4.25} arises due to the non-zero curvature of $F_{\sf vac}$. A similar term arises when squaring the Pauli operator from \eqref{4.19} and the Dirac operator from \eqref{4.23}. Note that in the limit $w^2\to\infty$, the operator $\Delta_6$ as well as $\Gamma_{\R^6}^{2}$ tend to the operator $\Delta_3$, i.e. $w^{-2}$ characterizes the strength of interaction of a particle with the field $\Av$.

For Clifford algebras on these spaces we have
\begin{equation}\label{4.26}
\mbox{Cl}^\C(7)= \mbox{Mat}(8, \C)\oplus\mbox{Mat}(8, \C)\ ,
\end{equation}
\begin{equation}\label{4.27}
\mbox{Cl}^\C(8)=\mbox{Mat}(16, \C)\ ,
\end{equation}
i.e. the representation of $\mbox{Cl}^\C(7)$ is reducible. Accordingly, on spaces $\R^{6,1}$ and $\R^{6,2}$ we have the operators
\begin{equation}\label{4.28}
\square_7 = -\dpar_0^2 + \Delta^{}_{\R^6}\und \Gamma^{}_{\R^{6,1}}=\Gamma^0\dpar_0+\Gamma^{}_{\R^{6}}\ ,
\end{equation}
\begin{equation}\label{4.29}
\square_8=\eta^{\mu\nu}\nabla_\mu\nabla_\nu + g^{}_{\mu +4\,\nu+4}\nabla^{\mu +4}\nabla^{\nu+4}\und\Gamma^{}_{\R^{6,2}}=\Gamma^\mu\nabla_\mu + \Gamma^{}_{\mu +4}\nabla^{\mu +4},
\end{equation}
where $x^0 =t$, $\mu , \nu =0,...,3$ and
\begin{equation}\label{4.30}
g^{}_{\mu +4\,\nu+4}=\frac{1}{w^4}\,\eta^{}_{\mu\nu}\ ,\quad\nabla_\mu =\dpar_\mu\und\nabla^{\mu +4}=
\frac{\dpar}{\dpar p_\mu}+ A^{\mu +4}J\ .
\end{equation}
Here $\Gamma^0$, $\Gamma^a$,$\Gamma_{a+3}$ are generators of the algebra \eqref{4.26} and $\Gamma^\mu $, $\Gamma_{\mu +4}$ are generators of the algebra  \eqref{4.27}, $A^{\mu +4}=-x^\mu$. Using these operators, we introduce equations
\begin{equation}\label{4.31}
(\square_7 - m^2)\Phi =0\und (\Gamma^{}_{\R^{6,1}}-m)\Psi=0\ ,
\end{equation}
\begin{equation}\label{4.32}
(\square_8 - m^2)\Phi =0\und (\Gamma^{}_{\R^{6,2}}-m)\Psi=0
\end{equation}
and consider their solutions.

\section{Extended Dirac equations on $\R^{1,1} \hra \R^{2,1}\hra T^*\R^{1,1}$}

\subsection{Preliminary remarks}

\noindent In the standard Dirac equation \eqref{4.23} there are no terms of interaction with the vacuum field  $\Av\in\ru(1)_{\sf v}$ since $\Av$ does not have components along the coordinate space $\R^{3,1}$. To see the effect of these $\Av$ on fermions we extend the Dirac equation to the equation \eqref{4.31} on the space $\R^{6,1}$ and to the equation \eqref{4.32} on $\R^{6,2}$ with the condition that spinors depend only on space-time coordinates $x^\mu\in\R^{3,1}$. 
When describing solutions to the corresponding Dirac equation with all gauge fields except the field $\Av$ turned off, the complicated form of various matrix differential operators makes it difficult to compare solutions with the standard case. We will construct solutions of equations for fermions $\Psi$, interacting with fields $\Av$.  The difference of these fermions $\Psi$ from free particles is that they interact with the vacuum field $\Av$, and they are no less real than free particles. Rather, on the contrary, free (bare) particles are a mathematical abstraction, as Bogoliubov and Shirkov insist \cite{BogShir}. 

For greater clarity, we will first consider solutions to the Dirac equations on space-time $\R^{1,1}$, compare them with those on $\R^{2,1}=\R\times T^*\R$ and  $\R^{2,2}=T^*\R^{1,1}$, and only then return to the case $T^*\R^{3,1}$. Recall that we consider spinors $\Psi$ as sections of the bundle $L_\C^+\oplus L_\C^-$ over $T^*\R^{1,1}$, i.e.
\begin{equation}\label{5.1} 
\Psi(x,t)=\psi_+v_+ + \psi_-v_-\ ,
\end{equation}
where $v_+$ is the basis in the quantum bundle $L_\C^+$, and $v_-$ is the basis in $L_\C^-$.

\subsection{Free particles in $\R^{1,1}$}

Consider  two-dimensional space-time $\R^{1,1}$ with the metric $\eta =(\eta_{AB})=\diag (1,-1)$, $A, B=0,1$. The Clifford 
algebra $\rCl(1,1)$ of this space has a real Majorana representation by matrices Mat(2,$\R$) and a complex Dirac representation of the algebra $\rCl^\C (2) = \rCl(1,1)\otimes \C$ by matrices Mat(2,$\C$). As generators of this algebra we take the matrices
\begin{equation}\label{5.2} 
\gamma^0 =\sigma_3\quad\mbox{and}\quad \gamma^1 =-\im\sigma_1\ .
\end{equation}
The Dirac equation for a free fermion of mass $m$ has the form
\begin{equation}\label{5.3}
(\im\gamma^A\dpar_A - m)\Psi = (\im\sigma_3\dpar_t +\sigma_1\dpar_1 - m)\Psi =0\ ,
\end{equation}
where $\dpar_A=\dpar/\dpar x^A , x^0 =t, x^1=x$. Positive frequency solution of equation \eqref{5.3} is
\begin{equation}\label{5.4}
\psi_+=\frac{1}{\sqrt{2\omega_p}}e^{-\im\omega_pt + \im px}u(p)\ ,\quad u(p)=\begin{pmatrix}\sqrt{\omega_p+m}\\[2pt]-\im\sqrt{\omega_p-m}\end{pmatrix}\ ,\quad \omega_p=\sqrt{p^2+m^2}\ .
\end{equation}
Negative frequency solution has the form
\begin{equation}\label{5.5}
\psi_-=\frac{1}{\sqrt{2\omega_p}}e^{\im\omega_pt - \im px}v(p)\ ,\quad v(p)=\begin{pmatrix}\im\sqrt{\omega_p-m}\\[2pt]
\sqrt{\omega_p+m}\end{pmatrix}\ ,\quad \omega_p=\sqrt{p^2+m^2}\ .
\end{equation}
Hence, we have the plane-wave solution
\begin{equation}\label{5.6}
\Psi = a_p\psi_+v_+ + b_p\psi_-v_-\quad\mbox{with}\quad\overline{\Psi}^\qv=\Psi^\+\gamma^0\otimes Q_{\sf v}=\overline{\Psi}Q_{\sf v}=a_p^\+\bar\psi_+v_+^\+ - b_p^\+\bar\psi_-v_-^\+
\end{equation}
for which
\begin{equation}\label{5.7}
\overline{\Psi}^\qv\Psi =\frac{m}{\omega_p}(a_p^\+ a_p + b_p^\+ b_p)>0\ .
\end{equation}
Here $a_p:=a(p)$ and $b_p:=b(p)$ are arbitrary complex-valued functions of momentum $p$ which become annihilation operators after the second quantization. From the solution  \eqref{5.6}, a general form of a wave packet solution is 
\begin{equation}\label{5.8}
\begin{split}
\Psi (x,t) &= \frac{1}{2\pi}\mathop{\int}_{-\infty}^{\infty}\dd p\, (a_p\psi_+v_++b_p\psi_-v_-)\\
&=\frac{1}{2\pi}\mathop{\int}_{-\infty}^{\infty}\dd p\frac{1}{\sqrt{2\omega_p}}\left (a_pu(p)e^{-\im\omega_pt+\im px}v_+ +
b_pv(p)e^{\im\omega_pt-\im px}v_-\right )\ .
\end{split}
\end{equation}
From \eqref{5.4}-\eqref{5.8} it is easy to deduce that energy of all solutions is positive.

For the quantum charge density for \eqref{5.6} we get
\begin{equation}\label{5.9}
\rho =\overline{\Psi}^{\qv}\gamma^0\Psi = a_p^\+a_p-b_p^\+b_p\ ,
\end{equation}
where $\rho\ge 0$  for ``particles" $a_p\psi_+$ and $\rho\le 0$ for ``antiparticles" $b_p\psi_-$. Note that
\begin{equation}\label{5.10}
\Psi = a_p\psi_+ v_+ + b_p\psi_-v_- = \frac{1}{\sqrt 2}
\begin{pmatrix}a_p\psi_+ + b_p\psi_-\\ -\im (a_p\psi_+ - b_p\psi_-)\end{pmatrix}\ ,
\end{equation}
and a return to usual discussion of solutions will occur if we discard the lower component in \eqref{5.10} of the $\C^2$-valued spinor and consider only upper component in \eqref{5.10}. 

The charge conjugated spinor $\Psi_c$ is introduced through the complex conjugation of the Dirac equation and multiplying it on the left by a certain matrix to reduce it to its original form. If gauge fields are turned on, then the form of the covariant 
derivative will change according to the change in charges.  In the case under consideration we have
\begin{equation}\label{5.11}
\Psi_c=\sigma_1\Psi^*=\sigma_1\psi^*_-v_+ + \sigma_1\psi^*_+v_-\ ,
\end{equation}
which implies that 
\begin{equation}\label{5.12}
(\psi_+)_c =\sigma_1\psi^*_-\quad\mbox{and}\quad(\psi_-)_c =\sigma_1\psi^*_+\ ,
\end{equation}
which is consisted with \eqref{5.4} and \eqref{5.5}.

\subsection{Dirac equation on $\R^{2,1}\subset T^*\R^{1,1}$}

\noindent Now we will consider a generalization of the Dirac equation from space $\R^{1,1}$ to phase space $T^*\R^{1,1}=\R^{1,1}\times \R^{1,1}=\R^{2,2}$ with the metric 
\begin{equation}\label{5.13}
g^{}_{\R^{2,2}}=- (\dd x^0)^2 + (\dd x^1)^2 +w_1^4\dd p_1^2 - w_0^4\dd p_0^2\ ,
\end{equation}
where $w_0$ and $w_1$ are length parameters (cf. \eqref{2.8}). As symplectic structure we consider
\begin{equation}\label{5.14}
\omega^{}_{\R^{2,2}}=\dd x^0\wedge \dd p_0 + \dd x^1\wedge \dd p_1\ .
\end{equation}
First we will consider solutions to the Dirac equation on space $\R^{2,1}\subset\R^{2,2}$ with coordinates $x^0, x^1$ and $p_1$, and then we will describe solutions on space $\R^{2,2}$ to compare and see the differences.

Note that the real Clifford algebras in three dimensions have Majorana representations
\begin{equation}\label{5.15}
\rCl(1,2)\cong\mbox{Mat}(2,\C)\und \rCl(2, 1)\cong\mbox{Mat}(2,\R)\oplus\mbox{Mat}(2,\R)\ ,
\end{equation}
and over the field $\C$ we have a reducible Dirac representation
\begin{equation}\label{5.16}
\rCl^\C(3)\cong \mbox{Mat}(2,\C)\oplus\mbox{Mat}(2,\C)\ .
\end{equation}
We will take the basis matrices in $\rCl(1,2)$ as
\begin{equation}\label{5.17}
\gamma^0=\sigma_3\ ,\ \ \gamma^1=-\im\sigma_1\ ,\ \ \gamma^2=-\frac{\im}{w^2}\,\sigma_2\ \ \Rightarrow\ \ \gamma^1\gamma^2=-\frac{\im}{w^2}\,\sigma_3\ ,
\end{equation}
From \eqref{5.17} we see that $\gamma^0$ and $\gamma^1\gamma^2$ are independent over $\R$  but equivalent over the field $\C$. Therefore, we should introduce $4\times 4$ matrices
\begin{equation}\label{5.18}
\Gamma^0{=}\begin{pmatrix}\gamma^0&0\\0&{-}\gamma^0\end{pmatrix},\ 
\Gamma^1{=}\begin{pmatrix}\gamma^1&0\\0&{-}\gamma^1\end{pmatrix},\ 
\Gamma_2{=}\begin{pmatrix}\gamma^2&0\\0&{-}\gamma^2\end{pmatrix} \Rightarrow
\Gamma^0\Gamma^1\Gamma_2{=}\im\Gamma_3{=}\frac{\im}{w^2}\begin{pmatrix}{-}\unit_2&0\\ 0&\unit_2\end{pmatrix},
\end{equation}
and we see that $\Gamma^1\Gamma_2$ and $\Gamma^0$ are not equivalent.

We use the notation
\begin{equation}\label{5.19}
\dpar_0{=}\frac{\dpar}{\dpar x^0}{=}\dpar_t,\ \ \dpar_1{=}\frac{\dpar}{\dpar x^1},\ \ \dpar^2{=}\frac{\dpar}{\dpar p_1},\ \
\nabla_0{=}\dpar_0, \ \ \nabla_1{=}\dpar_1,\ \ \nabla^2{=}\dpar^2 + A^2J ,\ \ 
A^2=-x^1 \ .
\end{equation}
With this notation the Dirac equation in $\R^{2,1}$ has the form
\begin{equation}\label{5.20}
(\im\Gamma^0\nabla_0 + \im\Gamma^1\nabla_1+\im\Gamma_2\nabla^2 - m)\Psi =0\ .
\end{equation}
Recall that $J$ is the generator of the structure group U(1)$_{\sf v}$ of the bundle $L_\C^+\oplus L^-_\C$ over $\R^{2,1}$ and $\Av$ in \eqref{5.19} is a connection on this bundle. The field $\Psi (x,t)$ in \eqref{5.20} is a section of the bundle $ L^+_\C\oplus L^-_\C$,
\begin{equation}\label{5.21}
\Psi =\Psi_+v_+ + \Psi_-v_-=\begin{pmatrix}\Psi_+^L\\\Psi_+^R\end{pmatrix}\otimes v_+ + 
\begin{pmatrix}\Psi_-^L\\\Psi_-^R\end{pmatrix}\otimes v_-
\end{equation}
and $\Psi^L$, $\Psi^R$ are the eigenvectors of the matrix $\Gamma_3$ from \eqref{5.18}.

For $\Psi^L_\pm$, $\Psi^R_\pm$, equations \eqref{5.20} are split into four equations
\begin{equation}\label{5.22}
(\im\sigma_3\dpar_t + \sigma_1\dpar_1 + w_1^{-2}\sigma_2\nabla^2_\pm - m)\Psi_\pm^L =0\ ,
\end{equation}
\begin{equation}\label{5.23}
(\im\sigma_3\dpar_t + \sigma_1\dpar_1 + w_1^{-2}\sigma_2\nabla^2_\pm + m)\Psi_\pm^R =0\ ,
\end{equation}
where $\nabla_\pm^2 = \dpar^2 \mp\im x^1$ and $\nabla_\pm^2 \Psi= \mp\im x^1\Psi$ since $\dpar^2\Psi =0$.

Equations \eqref{5.22} can be rewritten as
\begin{equation}\label{5.24}
\begin{pmatrix}\omega - m&-\frac{\sqrt 2}{w_1}\,a_1^\+\\ \frac{\sqrt 2}{w_1}\,a_1&-(\omega +m)\end{pmatrix}
\begin{pmatrix}\psi_+^1\\\psi_+^2\end{pmatrix}=0
\quad\mbox{for}\quad 
\Psi_+^L=e^{-\im\omega t}\begin{pmatrix}\psi_+^1\\\psi_+^2\end{pmatrix}\ ,
\end{equation}
where
\begin{equation}\label{5.25}
a_1:=\frac{w_1}{\sqrt 2}\left (\dpar_1+\frac{x^1}{w_1^2}\right )\ ,\quad 
a_1^\+:=-\frac{w_1}{\sqrt 2}\left (\dpar_1-\frac{x^1}{w_1^2}\right )\ ,\quad [a_1,a_1^\+]=1\ .
\end{equation}
Solutions of the Dirac equations \eqref{5.22} for $\Psi_+^L$ are 
\begin{equation}\label{5.26}
\Psi^L_{+,n}= \frac{e^{-\im\omega_nt}}{\sqrt{2\omega_n}}\begin{pmatrix}\sqrt{\omega_n+m} \,\,|n+1\rangle\\[3pt]
\sqrt{\omega_n-m} \,|n\rangle\end{pmatrix}\quad\Leftrightarrow\quad\psi_+=
\frac{e^{-\im\omega_pt}}{\sqrt{2\omega_p}}\,\begin{pmatrix}\sqrt{\omega_p+m}\,e^{\im px}\\[3pt]
-\sqrt{\omega_p-m}\,\im\, e^{\im px}\end{pmatrix}\ ,
\end{equation}
where
\begin{equation}\label{5.27}
\omega_n=\sqrt{2w_1^{-2}(n+1)+m^2}\quad\mbox{and}\quad\omega_p=\sqrt{p^2+m^2}\ .
\end{equation}
In \eqref{5.26} and \eqref{5.27} we wrote out for comparison the solutions of the Dirac equations for spinors $\Psi^L_{+,n}$ interacting with the field $\Av$ and the solution $\psi_+$ from \eqref{5.4} of the Dirac equation in $\R^{1,1}$ for free noninteracting spinors. We see that instead of momenta $p\in (-\infty , \infty)$, we get discrete numbers $n=0,1,...$ parametrizing oscillator-type solutions
\begin{equation}\label{5.28}
\Psi^L_{+,n}(x, t)= \frac{e^{-\im\omega_nt}}{\sqrt{2\omega_n}}
\begin{pmatrix}\sqrt{\omega_n+m}  \,\langle x|n+1\rangle\\[3pt]
\sqrt{\omega_n-m} \,\langle x|n\rangle\end{pmatrix}=
\frac{e^{-\im\omega_nt}}{\sqrt{2\omega_n}}
\begin{pmatrix}\sqrt{\omega_n+m}\,\,\psi_{n+1}(x)\\[3pt]
\sqrt{\omega_n-m}\,\,\psi_n(x)\end{pmatrix}\ ,
\end{equation}
where
\begin{equation}\label{5.29}
\psi_0(x)=\frac{1}{(\pi w_1^2)^{1/4}}\exp(-\frac{x^2}{2w_1^2})
\end{equation}
and $\psi_n(x)\sim H_n(\frac{x}{w_1})\,\psi_0(x)$, where $H_n$ are Hermitean polynomials.

Equations \eqref{5.22} for $\Psi_-^L$ have the form
\begin{equation}\label{5.30}
\begin{pmatrix}-(\omega+m)&\frac{\sqrt 2}{w_1}a_1\\[3pt]
-\frac{\sqrt 2}{w_1}a_1^\+&(\omega-m)\end{pmatrix}
\begin{pmatrix}\psi_-^1\\[3pt]
\psi_-^2\end{pmatrix}=0
\quad\mbox{for}\quad
\Psi_-^L=e^{\im\omega t}\,\begin{pmatrix}\psi_-^1\\[3pt]
\psi_-^2\end{pmatrix}\ ,
\end{equation}
and their solutions are
\begin{equation}\label{5.31}
\Psi^L_{-,n}= \frac{e^{\im\omega_nt}}{\sqrt{2\omega_n}}\begin{pmatrix}\sqrt{\omega_n-m} \,|n\rangle\\[3pt]
\sqrt{\omega_n+m} \,|n+1\rangle\end{pmatrix}=\left(\Psi_{+,n}^L\right )_c\ ,
\end{equation}
where $\Psi_c$ is defined in \eqref{5.11} and \eqref{5.12}. So, for $\Psi^L$ we obtain the general solution in the form
\begin{equation}\label{5.32}
\Psi^L=\mathop{\sum}_{n=0}^{\infty}\left(a_n\Psi_{+,n}^Lv_+ + b_n\Psi_{-,n}^Lv_-\right)\ ,
\end{equation}
which can be compared with \eqref{5.8}. We see that instead of a continuous set $(a_p, b_p, p\in\R )$ of functions parametrized by $p$, solutions \eqref{5.32} are parametrized by a discrete set $(a_n, b_n, n\in\Nbb )$ of complex numbers and the discrete set of energies \eqref{5.27}.

Note that from the Dirac oscillator equation \eqref{5.22}, which has the form
\begin{equation}\label{5.33}
\left(\im\sigma_3\dpar_t + \sigma_1\dpar_1 \mp \frac{\im x^1}{w_1^2}\,\sigma_2 - m\right)\Psi_\pm^L =0\ ,
\end{equation}
 follows the Klein-Gordon oscillator equation
\begin{equation}\label{5.34}
\left(-\dpar_t^2 + \dpar_x^2 - m^2 - \frac{ x^2}{w_1^4} - \frac{\sigma_3}{w_1^2}\right)\Psi_\pm^L =0\ ,
\end{equation}
with $x\equiv x^1$.  Note that solutions \eqref{5.32} are localized in space, unlike wave type solutions \eqref{5.8}. Solutions that are also localized in time will be written out bellow. After the second quantization of  \eqref{5.32} we will have $\{a_m, a_n^\+\}=\delta_{mn}$  instead of $\{a_p, a_{p^\prime}^\+\}=\delta(p-p^\prime)$ and similarly for $b_m$ and $b_p$.

For $\Psi_\pm^R$ all calculations are the same and we obtain solutions of the Dirac oscillator equations \eqref{5.23} in the form
\begin{equation}\label{5.35}
\begin{aligned}
\Psi^R_{+,n}&= \frac{e^{-\im\wt\omega_nt}}{\sqrt{2\wt\omega_n\vphantom{T^2_1}}}\begin{pmatrix}\sqrt{\wt\omega_n-\wt m\vphantom{T^2_1}} \,\,|n+1\rangle\\[3pt]
\sqrt{\wt\omega_n+\wt m\vphantom{T^2_1}} \,|n\rangle\end{pmatrix}\ ,\quad\wt\omega_n=\sqrt{2w_1^{-2}(n+1)+\wt m^2\vphantom{T^2_1}}\\
\Psi^R_{-,n}&= \frac{e^{\im\wt\omega_nt}}{\sqrt{2\wt\omega_n\vphantom{T^2_1}}}\begin{pmatrix}\sqrt{\wt\omega_n+\wt m\vphantom{T^2_1}} \,|n\rangle\\[3pt]
\sqrt{\wt\omega_n-\wt m\vphantom{T^2_1}}\, \,|n+1\rangle\end{pmatrix}=\left(\Psi^R_{+,n}\right)_c\ ,
\end{aligned}
\end{equation}
where we replaced $m$ in \eqref{5.23} with $\wt m$ since  \eqref{5.22} and \eqref{5.23} are independent. The general solution of the Dirac oscillator equation \eqref{5.23} is
\begin{equation}\label{5.36}
\Psi^R=\mathop{\sum}_{n=0}^{\infty}\left(c_n\Psi_{+,n}^R v_+ + d_n\Psi_{-,n}^R v_-\right)\ ,
\end{equation}
where $(c_n, d_n, n\in\Nbb)$ are independent of $(a_n, b_n, n\in\Nbb)$. All solutions obtained have positive energy, the quantum charge density is positive for $a_n, c_n$ and negative for $b_n, d_n$.

\subsection{Squeezed coherent states}

\noindent Note that in addition to solutions \eqref{5.26}-\eqref{5.28} with $n\ge 0$, equations \eqref{5.24} have a solution
\begin{equation}\label{5.37}
\Psi^L_{0,+}=C_0e^{-\im\omega_0t}\begin{pmatrix}\,|0\rangle\\0\end{pmatrix}\quad\mbox{with}\quad
\Psi^L_{0,+}(x_1, t)=C_0e^{-\im\omega_0t}\begin{pmatrix}\psi_0(x_1)\\0\end{pmatrix}
\end{equation}
different from $\Psi^L_{+,0}(x,t)$ in \eqref{5.28}. Here $C_0$ is a constant and $\psi_0(x_1)=\langle x_1|0\rangle$ is given in \eqref{5.29}.
Let us introduce an operator
\begin{equation}\label{5.38}
c_1=\frac{\sqrt 2}{w_1}\,a_1 =\nabla_1+\frac{\im}{w_1^2}\,\nabla_+^2=\dpar_1 + \frac{x^1}{w_1^2}
\end{equation}
which is a combination of covariant derivatives in the bundle $L_\C^+$ annihilating the state \eqref{5.37}. Using an automorphism of the bundle $L_\C^+$ given by the element $g\in\sU(1)_{\sf{v}}$, 
\begin{equation}\label{5.39}
g=e^{\im\vph}\quad\mbox{with}\quad \vph = x^1_{(0)}p_1 -p_1^{(0)} x^1\ ,
\end{equation}
we obtain a new connection $A_{\sf vac}^\vph$,
\begin{equation}\label{5.40}
A_{\sf vac}^\vph :\quad A^\vph_{x_1}=\im\dpar_{x_1}\vph = -\im p_1^{(0)}\ ,\quad 
A^\vph_{p_1}=A_{p_1}+\im\dpar_{p_1}\vph = -\im (x^1 - x^1_{(0)})\ ,
\end{equation}
that is not equivalent to the initial one, since the field $\Av$ is massive. In this case, we obtain
\begin{equation}\label{5.41}
c_1=\dpar_1+\frac{x^1}{w_1^2}\quad\mapsto\quad c_1^\vph =\dpar_1-\im p_1^{(0)}+\frac{(x^1-x^1_{(0)})}{w_1^2}
\end{equation}
and the new solution to the Dirac oscillator equation is a squeezed coherent state
\begin{equation}\label{5.42}
\Psi_0^{\rm{squ}}(x^1, t)=C_0\,\exp\left({-\frac{(x^1-x^1_{(0)})^2}{2w_1^2}-\im\omega_0t + \im p_1^{(0)}x^1}\right)\ ,\quad\omega_0=\sqrt{2w_1^{-2}+m^2\vphantom{T^2_1}},
\end{equation}
where $x^1_{(0)}$ is the center of the wave packet, $w_1$ is its width and $p_1^{(0)}$ is the expectation value of its momentum. 
In the limit $w_1^2\to\infty$, \eqref{5.42} becomes a plane-wave solution of free Dirac equation.

Note that \eqref{5.42} can also be obtained by acting on \eqref{5.37} by the operators
\begin{equation}\label{5.43}
D(\al)=e^{\al a_1^\+ - \al^*a_1}_{}\ ,\quad S(\rho)=e^{\sfrac12\rho(a_1^2 - a_1^{\+\,2})}_{} :\    \,|\al , \rho\rangle = D(\al )S(\rho) \,|0\rangle ,
\end{equation}
where the parameters $\rho$ and $\al =\al_1+\im\al_2$ can be expressed in terms of parameters $x^1_{(0)}, p_1^{(0)}$ and $w_1$. Here $\,|0\rangle$ is the vacuum state in \eqref{5.37}, $D(\al )$ is the displacement operator and $S(\rho)$ is the squeeze operator. The state \eqref{5.42} saturates the Heisenberg uncertainty relation $\Delta x\Delta p = \frac{\hbar}{2}$.

\subsection{Dirac equation on $T^*\R^{1,1}$}

\noindent Let us now consider the phase space $T^*\R^{1,1}=\R^{2,2}$ with metric \eqref{5.13} and symplectic form \eqref{5.14}.  The coordinate time $x^0$ is the time measured by a stationary clock in an inertial frame. The proper time $\tau$ of a particle is the time measured by a clock that moves with it and we can also introduce an evolution parameter $t$ different from $x^0$ and $\tau$, so that $x^0=x^0(t), \tau =\tau (t)$. Then we can promote $x^0$ to an operator and to see how this will change solutions of the Dirac oscillator equation on $\R^{2,1}$ with $\Av \ne 0$.

We consider the bundle $L_\C^+\oplus L_\C^-$ over $T^*\R^{1,1}$ with covariant derivative
\begin{equation}\label{5.44}
\nabla_0=\dpar_0\ ,\quad \nabla_1=\dpar_1\ ,\quad \nabla^2=\frac{\dpar}{\dpar p_1} - x^1J\ ,\quad
\nabla^3=\frac{\dpar}{\dpar p_0} - x^0J\ ,
\end{equation}
where
\begin{equation}\label{5.45}
J=\im\,(v_+ v_+^\+ - v_- v_-^\+)\quad\mbox{with}\quad v_\pm = \frac{1}{\sqrt 2}\begin{pmatrix}1\\\mp\im\end{pmatrix}
\end{equation}
is the generator of the group $\sU(1)_{\sf v}$ acting in the bundle $L_\C^+\oplus L_\C^-$. In \eqref{5.44} we see the components $x^0$ and $x^1$ of the vacuum connection $\Av$ on the quantum bundle $L_\C^+\oplus L_\C^-$. For sections $\psi =\psi_+v_+ + \psi_-v_-$ of this bundle depending only on $x^0$, $x^1$, we have
\begin{equation}\label{5.46}
\nabla^3\psi = -\im x^0\psi_+ v_+ +\im x^0\psi_-v_-\quad\mbox{and}\quad\nabla^2\psi = -\im x^1\psi_+ v_+ +\im x^1\psi_-v_-\ .
\end{equation}
For nonvanishing commutators of covariant derivatives \eqref{5.44} we have
\begin{equation}\label{5.47}
[\nabla_0, \nabla^3]=[\nabla_1, \nabla^2]=-J\ \Rightarrow\ [\nabla_0, \nabla^3]\psi_\pm=[\nabla_1, \nabla^2]\psi_\pm =\mp\im\psi_\pm\ ,
\end{equation}
and we can introduce operators
\begin{equation}\label{5.48}
\ph_0=-\im\dpar_0\ ,\quad\ph_1=-\im\dpar_1\ ,\quad\xh^0 = J \nabla^3=x^0+J\frac{\dpar}{\dpar p_0}\ ,\quad \xh^1 =J \nabla^2=x^1+J\frac{\dpar}{\dpar p_1}\ ,
\end{equation}
where the upper sign in \eqref{5.47} corresponds to $L^+_\C$, and lower sign there corresponds to the action on sections of $L_\C^-$.

As generators of Clifford algebra Cl(2,2)$\otimes\C$ we choose matrices
\begin{equation}\label{5.49}
\ga^0{=}\begin{pmatrix}0&\unit_2\\\unit_2&0\end{pmatrix},\
\ga^1{=}\begin{pmatrix}0&\sigma_1\\-\sigma_1&0\end{pmatrix},\
\ga_2{=}\frac{1}{w_1^2}\begin{pmatrix}0&\sigma_2\\-\sigma_2&0\end{pmatrix}\und
\ga_3{=}\frac{1}{w_0^2}\begin{pmatrix}0&\im\sigma_3\\-\im\sigma_3&0\end{pmatrix}.
\end{equation}
We consider spinors $\Psi$ with values in the bundle $L_\C^+\oplus L_\C^-$,
\begin{equation}\label{5.50}
\Psi =\Psi_+ v_+ + \Psi_- v_-
\end{equation}
and their conjugate $\overline{\Psi}^{\qv}:=\Psi^\+\Gamma\otimes Q_{\sf v}$, where 
\begin{equation}\label{5.51}
\Gamma :=\frac{\im}{w_1^2}\ga^1\ga_2 = \frac{1}{w_1^2}\begin{pmatrix}\sigma_3&0\\0&\sigma_3\end{pmatrix}\ ,\quad
Q_{\sf v}=-\im J=v_+v_+^\+ - v_-v_-^\+=-\sigma_2\ ,
\end{equation}
and the scalar product is $\overline{\Psi}^\qv\Psi$.

The Dirac equation for $\Psi$ on $T^*\R^{1,1}$ has the form
\begin{equation}\label{5.52}
(\im\ga^0\nabla_0+\im\ga^1\nabla_1+\im\ga_2\nabla^2+\im\ga_3\nabla^3 - m)\Psi =0\ ,
\end{equation}
where
\begin{equation}\label{5.53}
\nabla_0{=}\dpar_0,\ \nabla_1{=}\dpar_1,\ \nabla^2{=}\dpar^2{-}x^1J
\und
\nabla^3{=}\dpar^3{-}x^0J\ .
\end{equation}
 From \eqref{5.53} we obtain
\begin{equation}\label{5.54}
\nabla^2\Psi_\pm=\mp\frac{\im x^1}{w_1^2}\Psi_\pm\quad\mbox{and}\quad\nabla^3\Psi_\pm=\mp\frac{\im x^0}{w_0^2}\Psi_\pm\ .
\end{equation}

\subsection{Quantum time and fermions}

In \eqref{5.44} and \eqref{5.48} we introduced the quantum time operator $\xh^0$ acting on the wave functions according the formulas  \eqref{5.46} and \eqref{5.54}. Recall that $\dpar_{p_0}\Psi =\dpar_{p_1}\Psi =0$ and operators \eqref{5.44} are combined into operators of creation and annihilation,
\begin{equation}\label{5.55}
c_0=\frac{\sqrt 2}{w_0}\,a_0=\dpar_0 + \frac{x^0}{w_0^2}\ ,\quad
c_0^\+=\frac{\sqrt 2}{w_0}\,a_0^\+=-\left(\dpar_0 - \frac{x^0}{w_0^2}\right )\ ,
\end{equation}
\begin{equation}\label{5.56}
c_1=\frac{\sqrt 2}{w_1}\,a_1=\dpar_1 + \frac{x^1}{w_1^2}\quad\mbox{and}\quad
c_1^\+=\frac{\sqrt 2}{w_1}\,a_1^\+=-\left(\dpar_1 - \frac{x^1}{w_1^2}\right )\ .
\end{equation}
Using these operators, equation \eqref{5.52} is rewritten as
\begin{equation}\label{5.57}
\begin{pmatrix}c_0&-c_1^\+\\c_1&-c_0^\+\end{pmatrix}\psi_+^2 + \im\,m\,\psi_+^1=0\ ,\quad 
\begin{pmatrix}-c_0^\+&c_1^\+\\-c_1&c_0\end{pmatrix}\psi_+^1 + \im\,m\,\psi_+^2=0\ ,
\end{equation}
\begin{equation}\label{5.58}
\begin{pmatrix}-c_0^\+&c_1\\-c_1^\+&c_0\end{pmatrix}\psi_-^2 + \im\,m\,\psi_-^1=0\ ,\quad 
\begin{pmatrix}c_0&-c_1\\c_1^\+&-c_0^\+\end{pmatrix}\psi_-^1 + \im\,m\,\psi_-^2=0\ .
\end{equation}
Here we used the substutution 
\begin{equation}\label{5.59}
\Psi_+=\begin{pmatrix}\psi_+^1\\\psi_+^2\end{pmatrix}\in\C^4\quad\mbox{and}\quad
\Psi_-=\begin{pmatrix}\psi_-^1\\\psi_-^2\end{pmatrix}\in\C^4
\end{equation}
in \eqref{5.50} and \eqref{5.52}.

As solutions of equations \eqref{5.57}, we get
\begin{equation}\label{5.60}
\Psi_+(n_0,n_1):\quad \psi_+^1{=}\begin{pmatrix}\al_+ |n_0,n_1+1\rangle\\
\al_- |n_0+1,n_1\rangle \end{pmatrix},\
\psi_+^2{=}\frac{\im}{m}\begin{pmatrix}(-\al^{}_+\omega_{n_0}+\al^{}_-\omega_{n_1})\,|n_0+1, n_1+1\rangle\\
(\al^{}_-\omega_{n_0}-\al^{}_+\omega_{n_1})\,|n_0, n_1\rangle \end{pmatrix} ,
\end{equation}
where
\begin{equation}\label{5.61}
 \al_\pm =\sqrt{\omega_{n_0}\pm m},\ \omega_{n_0}=\sqrt{\omega_{n_1}^2+m^2\vphantom{T^2_1}}=\sqrt{2w_0^{-2}(n_0+1)\vphantom{T^2_1}},\ 
\omega_{n_1}^2=2w_1^{-2}(n_1+1)\ .
\end{equation}
Note that here we can change the operator $c_1$ as in \eqref{5.41} and similarly
\begin{equation}\label{5.62}
c_0=\dpar_0+\frac{x^0}{w_0^2}\quad\mapsto\quad
c_0^\vph =\dpar_0-\im p_0^{(0)}+\frac{(x^0-x^0_{(0)})}{w_0^2}\ ,
\end{equation}
and write the ground state as
\begin{equation}\label{5.63}
\psi_{0,0}(x^0, x^1)=C_0(w_0) C_1(w_1)e^{-\frac{(x^0-x^0_{(0)})^2}{2w_0^2}-\frac{(x^1-x^1_{(0)})^2}{2w_1^2}+\im p_0^{(0)}x^0+\im p_1^{(0)}x^1}\sim\langle x\,|n_0=0, n_1=0\rangle\ ,
\end{equation}
where $ C_0(w_0) C_1(w_1)$ is a normalization constant. We see that obtained solutions are localized both in space and time.

Depending on the choice of normalization coefficients in \eqref{5.63} we have various limiting cases:
\begin{equation}\label{5.64}
C_1(w_1)=1:\quad \mathop{\lim}_{w_1\to\infty}e^{-\frac{(x^1-x^1_{(0)})^2}{2w_1^2}+\im p_1^{(0)}x^1}= e^{\im p_1^{(0)}x^1},
\end{equation}
\begin{equation}\label{5.65}
C_1(w_1)=\frac{1}{\sqrt{2\pi w_1^2}}:\quad \mathop{\lim}_{w_1\to 0}\frac{1}{\sqrt{2\pi w_1^2}}\,e^{-\frac{(x^1-x^1_{(0)})^2}{2w_1^2}+\im p_1^{(0)}x^1}= \delta (x^1-x^1_{(0)})\,e^{\im p_1^{(0)}x^1}\ .
\end{equation}
Similarly we have
\begin{equation}\label{5.66}
C_0(w_0)e^{-\frac{(x^0-x^0_{(0)})^2}{2w_0^2}+\im p_0^{(0)}x^0}\ \longrightarrow\ \left\{
\begin{array}{ll}e^{\im p_0^{(0)}x^0} &\mbox{for}\ w_0\to\infty\\ \delta (x^0-x^0_{(0)})\,e^{\im p_0^{(0)}x^0}&\mbox{for}\ w_0\to\ 0\end{array}\right.
\end{equation}
i.e. we can have localization either in energy or in time.

Solutions of equations \eqref{5.58} have the form
\begin{equation}\label{5.67}
\Psi_-(n_0,n_1):\quad \psi_-^1{=}\begin{pmatrix}\al_- \,|n_0+1, n_1\rangle\\\al_+\,|n_0, n_1+1\rangle\end{pmatrix},\ 
\psi_-^2{=}\frac{\im}{m}\begin{pmatrix}(\al_-\omega_{n_0} - \al_+\omega_{n_1}) \,|n_0, n_1\rangle\\(-\al_+\omega_{n_0} + \al_-\omega_{n_1})\,|n_0+1, n_1+1\rangle\end{pmatrix}
\end{equation}
with the same formulae \eqref{5.61}. It is easy to verify that $\Psi_-$ written out in \eqref{5.67} is charge conjugate to $\Psi_+$ from \eqref{5.60},
\begin{equation}\label{5.68}
\Psi_-= \begin{pmatrix}\psi_-^1\\\psi_-^2\end{pmatrix}=C\Psi_+^*=-\ga^0\ga^1\Psi_+^*=
\begin{pmatrix}\sigma_1&0\\0&-\sigma_1\end{pmatrix}\begin{pmatrix}(\psi_+^1)^*\\(\psi_+^2)^*\end{pmatrix}\ ,
\end{equation}
where ``$*$" means complex conjugation. The general solution of the Dirac equation \eqref{5.52} is 
\begin{equation}\label{5.69}
\Psi=\mathop{\sum}_{n_0=0}^{\infty}\mathop{\sum}_{n_1=0}^{\infty}
\left(a_{n_0n_1}\Psi_{+}(n_0,n_1) v_+ + b_{n_0n_1}\Psi_{-}(n_0,n_1)v_-\right)\ ,
\end{equation}
where $a_{n_0n_1}$ and $b_{n_0n_1}$ are complex numbers, $n_0, n_1\in \mathbb N$.

\section{Extended Dirac equations on $\R^{3,1} \hra \R^{6,1}\hra T^*\R^{3,1}$}

\subsection{Gamma matrices}

\noindent Having completed the consideration of the two-dimensional case with a comparison of fermions 
(``coupled" or ``virtual") interacting with the vacuum field $\Av$  and fermions (``free" or ``bare") not interacting with any field, we move on to the four-dimensional case. Let us consider the space $\R^{6,1}\subset T^*\R^{3,1}=\R^{6,2}$ with the metric 
\begin{equation}\label{6.1}
g^{}_{\R^{6,1}} = -(\dd x^0)^2 + g^{}_{\R^{6}}\ ,
\end{equation}
where $g^{}_{\R^{6}}$ is the metric on the phase space $T^*\R^{3}$ given in \eqref{2.8}. Minkowski space $\R^{3,1}$ with the metric
\begin{equation}\label{6.2}
g^{}_{\R^{3,1}}=\eta_{\mu\nu}\dd x^\mu \dd x^\nu\ ,\quad \eta =(\eta_{\mu\nu})=\diag(-1, 1, 1, 1)
\end{equation}
is a subspace in $\R^{6,1}$. Clifford algebras for spaces $\R^{3,1}$ and $\R^{1,3}$ have matrix representation 
\begin{equation}\label{6.3}
\rCl(3,1)\cong \mbox{Mat} (4, \R)\quad\mbox{and}\quad \rCl(1,3)\cong \mbox{Mat} (2, \Hbb)\ ,
\end{equation}
where $\Hbb$ is the associative algebra of quaternions. Hence, spinors in the spaces $\R^{3,1}$ and $\R^{1,3}$ will be columns $\R^4$ and $\Hbb^2$, respectively, and these will be Majorana spinors. One can identify $\Hbb$ with $\C^2$ and then Mat(2,$\Hbb$) will be embedded into Mat(4,$\C$) as a subalgebra defined by some reality conditions. Instead, Dirac considered the complexified Cliford algebra,
\begin{equation}\label{6.4}
\rCl^\C(4):= \rCl(3,1)\otimes\C = \rCl(1,3)\otimes\C \cong \mbox{Mat}(4, \C)
\end{equation}
for which spinors $\Psi$ are $\C^4$-valued.

We choose generators of the algebra $\rCl^\C(4)$ as matrices
\begin{equation}\label{6.5}
\ga^0=\im\begin{pmatrix}\unit_2&0\\0&-\unit_2\end{pmatrix},\quad \ga^a=\begin{pmatrix}0&\im\sigma_a\\-\im\sigma_a&0\end{pmatrix},\quad \im\ga^0\ga^1\ga^2\ga^3 =\begin{pmatrix}0&\unit_2\\\unit_2&0\end{pmatrix}=:\ga^5\ ,
\end{equation}
where $\sigma_a$ are Pauli matrices. Generators \eqref{6.5} satisfy the anticommutation relations
\begin{equation}\label{6.6}
\{\ga^\mu, \ga^\nu\}=\ga^\mu\ga^\nu + \ga^\nu\ga^\mu =2\eta^{\mu\nu}\unit_4\quad\mbox{and}\quad \{\ga^\mu, \ga^5\}=0\ .
\end{equation}
For $\R^{6,1}$ and $\R^{6,2}$ we have
\begin{equation}\label{6.7}
\rCl^\C(7)\cong \mbox{Mat}(8, \C)\oplus\mbox{Mat}(8, \C)\quad\mbox{and}\quad \rCl^\C(8)\cong \mbox{Mat}(16, \C)\ ,
\end{equation}
analogously to the cases $\R^{2,1}$ and $\R^{2,2}$. We choose generators of $\rCl^\C(7)$ in the form 
\begin{equation}\label{6.8}
\wt\Gamma^0=\begin{pmatrix}\Gamma^0&0\\0&-\Gamma^0\end{pmatrix}, \quad 
\wt\Gamma^a=\begin{pmatrix}\Gamma^a&0\\0&-\Gamma^a\end{pmatrix}\und
\wt\Gamma_{a+3}=\begin{pmatrix}\Gamma_{a+3}&0\\0&-\Gamma_{a+3}\end{pmatrix},\ a=1,2,3,
\end{equation}
where $\Gamma^a, \Gamma_{a+3}\in\mbox{Mat}(8, \C)$ are generators of the algebra $\rCl^\C(6)$  introduced in   \eqref{4.11} and we will choose them in the form
\begin{equation}\label{6.9}
\Gamma^0{=}\ga^0\otimes\sigma_3,\ 
\Gamma^1{=}\ga^1\otimes\sigma_3,\ 
\Gamma^2{=}\ga^3\otimes\sigma_3,\ 
\Gamma^3{=}\ga^2\otimes\sigma_3,\ 
\Gamma_4{=}\frac{1}{w^2}\unit_4\otimes\sigma_2,\ 
\Gamma_5{=}\frac{1}{w^2}\,\ga^5\otimes\sigma_3,\ 
\Gamma_6{=}\frac{1}{w^2}\unit_4\otimes\sigma_1.
\end{equation}
We have ${\overline{\wt\Psi}}:=\wt\Psi^\+\hat\Gamma^0$ for $\wt\Psi\in\C^8\oplus\C^8$, $\hat\Gamma^0:=-\im\wt\Gamma^0$.

\subsection{Dirac equation on $\R^{6,1}$}

\noindent Over the space $\R^{6,1}$ there is given the quantum bundle $L_\C^+\oplus L_\C^-$ described in detail in the previous sections, and we consider spinors $\wt\Psi$ having an additional index of the two-dimensional space of fibres of the bundle $L_\C^+\oplus L_\C^-$,
\begin{equation}\label{6.10}
\wt\Psi = \begin{pmatrix}\Psi\\\Phi\end{pmatrix}=\begin{pmatrix}\Psi_+v_++\Psi_-v_-\\\Phi_+v_++\Phi_-v_-\end{pmatrix}\in
\begin{pmatrix}\C^{16}\\\C^{16}\end{pmatrix}\ .
\end{equation}
 The bundle $L_\C^+\oplus L_\C^-$ is endowed with the metric, quantum connection $\Av$ and covariant derivatives $\wt\nabla_\mu$, $\wt\nabla_{a+3}$ described in the previous sections. The Dirac equation for spinors $\wt\Psi$ from \eqref{6.10} has the form 
\begin{equation}\label{6.11}
(\wt\Gamma^0\dpar_0 + \wt\Gamma^a\nabla_a +\wt\Gamma_{a+3}\nabla^{a+3} - \wt m)\wt\Psi =0\quad\mbox{for}\quad \wt m=\diag (m\unit_{16},\ m^\prime\unit_{16})\ .
\end{equation}
Here the covariant derivatives are of the form
\begin{equation}\label{6.12}
\nabla_a=\dpar_a\ ,\quad \nabla^{b+3}=\dpar^{b+3}-x^bJ\ ,\quad
\dpar^{b+3}=\frac{\dpar}{\dpar p_b}\ ,
\end{equation}
so for commutators we have
\begin{equation}\label{6.13}
[\nabla_a, \nabla^{b+3}_+]=-\im\delta_a^b\ \mbox{on}\ L_\C^+\quad\mbox{and}\quad
[\nabla_a, \nabla^{b+3}_-]=\im\delta_a^b\ \mbox{on}\ L_\C^-\ .
\end{equation}
Note that the equations for $\Psi$ and $\Phi$ in \eqref{6.10} are independent, so we will consider only the equation for $\Psi$, for $\Phi$ everything is the same. From \eqref{6.10}-\eqref{6.12} we obtain equations for $\Psi_\pm$,
\begin{equation}\label{6.14}
(\Gamma^0\dpar_0 + \Gamma^a\dpar_a + \Gamma_{a+3}\nabla_\pm^{a+3} - m)\Psi_\pm =0\ ,
\end{equation}
from which there follow the Klein-Gordon oscillator equations
\begin{equation}\label{6.15}
\left(-\dpar_0^2 + \delta^{ab}\dpar_a\dpar_b - m^2 - \frac{1}{w^4}\,\delta_{ab}x^ax^b +  [\Gamma^a, \Gamma_{a+3}]\right)\Psi_\pm =0
\end{equation}
for components of spinors $\Psi_\pm\in\C^8$.

\subsection{Ladder operators}

\noindent For greater generality, we replace $w^2\to (w_1^2, w_2^2, w_3^2)$ (anisotropic case) and introduce the operators
\begin{equation}\label{6.16}
\begin{split}
c_1=\dpar_1 + \frac{x^1}{w_1^2}=:\frac{\sqrt 2}{w_1}\,a_1\ ,\quad&c_1^\+=-(\dpar_1 - \frac{x^1}{w_1^2})=:\frac{\sqrt 2}{w_1}\,a_1^\+\ ,\\
c_2=\dpar_2 + \frac{x^2}{w_2^2}=:\frac{\sqrt 2}{w_2}\,a_2\ ,\quad&c_2^\+=-(\dpar_2 - \frac{x^2}{w_2^2})=:\frac{\sqrt 2}{w_2}\,a_2^\+\ ,\\
c_3=\dpar_3 + \frac{x^3}{w_3^2}=:\frac{\sqrt 2}{w_3}\,a_3\ ,\quad&c_3^\+=-(\dpar_3 - \frac{x^3}{w_3^2})
=:\frac{\sqrt 2}{w_3}\,a_3^\+\ .
\end{split}
\end{equation}
Note that 
\begin{equation}\label{6.17}
[c_a, c_b^\+]=\frac{2}{w_a^2}\,\delta_{ab}\quad\mbox{and}\quad [a_b, a_c^\+]=\delta_{bc}
\end{equation}
and when $w_a^2\to\infty$ the first commutators will go to zero, but the second ones will not.

After some calculations we obtain 
\begin{equation}\label{6.18}
\Gamma_+:=\Gamma^a\dpar_a + \Gamma_{a+3}\nabla_+^{a+3} = -\im\begin{pmatrix}
0&0&c_2^\+&c_1^\+&c_3&0&0&0\\
0&0&-c_1&c_2&0&c_3&0&0\\
c_2&-c_1^\+&0&0&0&0&c_3&0\\
c_1&c_2^\+&0&0&0&0&0&c_3\\
c_3^\+&0&0&0&0&0&-c_2^\+&-c_1^\+\\
0&c_3^\+&0&0&0&0&c_1&-c_2\\
0&0&c_3^\+&0&-c_2&c_1^\+&0&0\\
0&0&0&c_3^\+&-c_1&-c_2^\+&0&0
\end{pmatrix}
\end{equation}
Now we set $\Psi_+=e^{-\im\omega t}\psi$ and obtain the equation
\begin{equation}\label{6.19}
(\im\omega \Gamma^0 + m -\Gamma_+)\psi =0\ ,
\end{equation}
where $\Gamma_+$ is given in \eqref{6.18}.

\subsection{Solutions}

\noindent Solutions of equation \eqref{6.19} are
\begin{equation}\label{6.20}
\begin{pmatrix}\psi_1\\\psi_2\\\psi_7\\\psi_8\end{pmatrix}{=}
\begin{pmatrix}\al_1|n_1+1, n_2+1, n_3\rangle\\\al_2|n_1, n_2, n_3\rangle\\ \al_3|n_1+1, n_2, n_3+1\rangle\\ \al_4|n_1, n_2+1, n_3+1\rangle \end{pmatrix},\
\begin{pmatrix}\psi_3\\\psi_4\\\psi_5\\\psi_6\end{pmatrix}{=}\frac{-\im}{(\omega_{n_1n_2n_3}+m)}
\begin{pmatrix}c_2\psi_1-c^\+_1\psi_2+c_3\psi_7\\ c_1\psi_1+c^\+_2\psi_2+c_3\psi_8\\
c_3^\+\psi_1-c^\+_2\psi_7-c_1^\+\psi_8\\
c_3^\+\psi_2+c_1\psi_7-c_2\psi_8\\\end{pmatrix}\ ,
\end{equation}
where $\al_1, \al_2, \al_3$ and $\al_4$ are arbitrary constants and
\begin{equation}\label{6.21}
\omega_{n_1n_2n_3}=\bigl(2w_1^{-2}(n_1+1) + 2w_2^{-2}(n_2+1) +2w_3^{-2}(n_3+1) + m^2\bigr)^\sfrac12\ .
\end{equation}
We do not write out the explicit form of the components $\psi_3, \psi_4, \psi_5, \psi_6$; they are easily calculated using \eqref{6.16} and \eqref{6.20}. 

It is easy to show that the charge conjugation operator for equation \eqref{6.19} is given by the matrix
\begin{equation}\label{6.22}
C=w^6\Gamma_4\Gamma_5\Gamma_6 = \begin{pmatrix}\im\sigma_2&0\\0&-\im\sigma_2\end{pmatrix}\otimes\sigma_1=
\begin{pmatrix}
0&0&\im\sigma_2&0\\
0&0&0&-\im\sigma_2\\
\im\sigma_2&0&0&0\\
0&-\im\sigma_2&0&0
\end{pmatrix}
\end{equation}
and we have solutions of the Dirac oscillator equation \eqref{6.11} for $\Psi\in \C^{16}$ in the form
\begin{equation}\label{6.23}
\Psi_{(n)}=\Psi_+^{(n)}v_+ + \Psi_-^{(n)}v_-\quad\mbox{with}\quad
\Psi_-^{(n)}=(\Psi_+^{(n)})_c= C(\Psi_+^{(n)})^*\ ,
\end{equation}
where $\Psi_+^{(n)}=e^{-\im\omega_{(n)}t}\psi_{(n)}$ is given in \eqref{6.20}, for $(n)=(n_1 n_2 n_3)$.

From the viewpoint of Minkowski space, equations \eqref{6.14} with solutions \eqref{6.23} describe two particles $(\qv =1)$ and two antiparticles 
$(\qv =-1)$. The equations for $\Phi$ from \eqref{6.10} are solved in exactly the same way, and also describe two particles and two antiparticles. The energy of all solutions is positive and $\overline{\wt\Psi}^\qv\wt\Psi = \wt\Psi^\+\hat\Gamma^0\otimes Q_{\sf v}\wt\Psi$ for $\wt\Psi$ from \eqref{6.10}.  If we lift the Dirac equation with $\Av$ from $\R^{6,1}\subset T^*\R^{3,1}$ to $T^*\R^{3,1}$, then the results will be similar to those discussed in \eqref{5.55}-\eqref{5.68} for $T^*\R^{1,1}$, the solutions will be bound states localized not only in space, but also in time. In fact, these solutions describe the supersymmetric Klein-Gordon oscillator. In \cite{PopovLMP}, we described them in terms of Grassmann variables for spinors and Bergman spaces of holomorphic functions.

\bigskip

\noindent 
{\bf\large Acknowledgments}

\noindent
I am grateful to Tatiana Ivanova for useful remarks.

\end{document}